\documentclass[prb,twocolumn,showpacs]{revtex4}
\usepackage[dvips]{graphicx}

\newcommand{\ignore}[1]{}
\newcommand{\beq}{\begin{equation}}
\newcommand{\eeq}{\end{equation}}

\begin{document}

\title{Realization of the mean-field universality class in spin-crossover materials}
\author{Seiji Miyashita$^{1,4}$, Yusuk\'e Konishi$^{1,4}$, Masamichi Nishino$^{2,4}$, Hiroko Tokoro$^{1}$, 
Per Arne Rikvold$^{3}$}
\affiliation{
$^{1}${\it Department of Physics, Graduate School of Science,
The University of Tokyo, 7-3-1 Hongo, Bunkyo-Ku, Tokyo 113-8656, Japan}  \\
$^{2}${\it Computational Materials Science Center, National Institute
for Materials Science, Tsukuba, Ibaraki 305-0047, Japan} \\
$^{3}$ {\it Department of Physics, Florida State University, Tallahassee, FL 32306-4350, USA} \\
$^{4}${\it CREST, JST, 4-1-8 Honcho Kawaguchi, Saitama 332-0012, Japan}
}
\date{\today}

\begin{abstract}

In spin-crossover materials, the volume of a molecule changes 
depending on whether it is in the high-spin (HS) or
low-spin (LS) state. This change causes distortion of the lattice. 
Elastic interactions among these 
distortions play an important role for the cooperative 
properties of spin-transition phenomena.
We find that the critical behavior caused by this elastic interaction 
belongs to the mean-field universality class,
in which the critical exponents for the spontaneous magnetization and 
the susceptibility are $\beta = 1/2$ and $\gamma = 1$, respectively.
Furthermore, the spin-spin correlation function 
is a constant at long distances, and it does not show
an exponential decay in contrast to short-range models.
The value of the correlation function
at long distances shows different size-dependences:
$O(1/N)$, $O(1/\sqrt{N})$, and constant for temperatures
above, at, and below the critical temperature, respectively.
The model does not exhibit clusters, even near the critical point. 
We also found that cluster growth is suppressed in the present model and 
that there is no critical opalescence in the coexistence region. 
During the relaxation process 
from a metastable state at the end of a hysteresis loop,
nucleation phenomena are not observed, and spatially uniform configurations are maintained during the change
of the fraction of HS and LS. 
These characteristics of the mean-field model are expected to be found not only
in spin-crossover materials, but also generally in systems 
where elastic distortion mediates the interaction among local states. 
\end{abstract}

\pacs{75.30.Wx 75.50.Xx 75.60.-d 64.60.-i}

\maketitle

\section{introduction}

Spin-crossover (SC) 
materials consist of local units (molecules), each of which has 
two different spin states, i.e., 
the low-spin (LS) and high-spin (HS) states.
The LS state is energetically 
favorable and dominates at low temperatures, while the
HS state dominates at high temperatures because it is entropically favorable. 
The transition between the LS and HS states is also induced 
by changes of the pressure, magnetic field, light-irradiation, 
etc.\cite{Gutlich,Decurtins,Kahn,Letard2,Hauser,Real,Shimamoto}
When interactions between molecules are weak, 
the HS fraction changes smoothly with temperature.
However, when the interactions become strong, the system exhibits cooperative 
phenomena.\cite{Sorai1} 
The change in the HS fraction becomes sharper with increasing interaction.
When the strength of the interaction exceeds a critical value, 
the change becomes discontinuous. 
In order to control electronic and magnetic properties of SC compounds, 
it is important to understand the bistable nature of such molecular solids.

\begin{figure}[t]
\centerline{\includegraphics[clip,width=6.5cm]{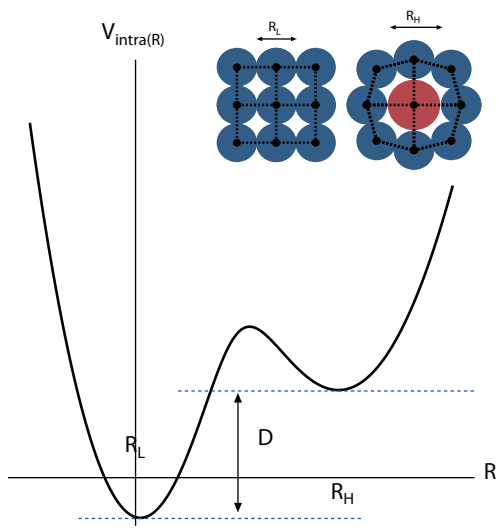} }
\caption{(Color online) 
Schematic picture of the energy structure of a molecule.
The left (right) minimum corresponds to the LS (HS) state.
In the inset, schematic pictures of a lattice of LS molecules (left), 
and the distortion caused by a HS  
molecule in a lattice of LS molecules (right) are illustrated.
}
\label{Fig_Vintra}
\end{figure}

As an important ingredient of the spin-crossover transition, 
we need two important characteristics of the system.
One of them is the structure of the intra-molecule Hamiltonian. 
At each molecule, we set an energy difference between the states $D(>0)$ 
(see Fig.~\ref{Fig_Vintra}) 
and different degeneracies of the states: $g_{\rm HS}$ and $g_{\rm LS}$ 
for the HS state and the LS state, respectively.
We express the spin state at the $i$-th site by $s_i$ 
which takes $-1$ for LS and $+1$ for HS.
The intra-molecule (on-site) interaction is expressed by
\begin{equation}
{H}_0={1\over 2}D\sum_is_i.
\end{equation}
If we take into account the effect of the degeneracy as a temperature dependent
field, we can use an effective Hamiltonian with non-degenerate
variables $\sigma_i=\pm 1$:
\begin{equation}
{\cal H}_{\rm eff} = \frac{1}{2}\sum_{i} (D - k_{\rm B} T \ln g ) \sigma_i,
\label{WP0}
\end{equation}
where $g=g_{\rm HS}/g_{\rm LS}$
denotes the degeneracy ratio between the HS and LS states.

The other important characteristic is the intermolecular interaction. 
For the cooperative property in the SC transition, 
until recently a short-range Ising-type
interaction has been adopted in the so-called Wajnflasz-Pick 
(WP) model:\cite{Wajnflasz2} 
\begin{equation}
{\cal H} =- J\sum_{\langle i,j\rangle }\sigma_{i} 
\sigma_{j} + \frac{1}{2}\sum_{i} (D - k_{\rm B} T \ln g ) \sigma_i.
\label{WP}
\end{equation} 
This type of model has successfully explained various aspects of the ordering 
processes.\cite{Bousseksou1,Kamel1,Nishino1,Nishino_dynamical,Miya2}
However, 
the origin of the interactions between the spin states has remained unclear.
There are various plausible origins of the interaction.

As a possible interaction mechanism, 
the importance of elastic interactions has been pointed 
out.\cite{Zimmermann,Kambara,Adler,Willenbacher,Tchougreeff,Spiering,Nasser,kbo} 
The elastic constants may depend on the neighboring spin states.
This dependence causes an effective interaction between the spin states.  
This effect of the elastic constants was investigated 
in a one-dimensional (1D) two-level model\cite{Nasser,KBO2} and also 
in a 1D vibronic coupling model.\cite{kbo}
In these one-dimensional versions of the model, 
the elastic interactions can be traced out locally, leading to an exact
mapping onto a 1D Ising ferromagnet, so that there is  
no phase transition at nonzero temperatures.\cite{kbo,KBO2} 

In higher spatial dimensions, as depicted in the inset in 
Fig.~\ref{Fig_Vintra}, 
the volume change of a molecule causes a distortion of the lattice. 
Elastic interactions mediate the effect of this distortion over long
distances. 
Therefore, in higher dimensions, the elastic interactions cause intrinsically 
different effects than in one dimension. 
We denote this long-range 
interaction by ${\cal H}_{\rm elastic}(\{\sigma_i\})$.
We do not know the explicit form of this interaction. (But see discussion
in Appendix B). 
However, we recently demonstrated that this type of 
elastic interaction can induce a phase transition in
spin-crossover systems.\cite{Hawaii,Nishino2007,Konishi2007}
This elastic interaction model is a kind of 
compressible Ising model,\cite{Fisher1968} and
similar models have been studied for binary alloys.\cite{Dunweg,Laradji,Zhu}

Because the interaction originating
from the elastic distortions is qualitatively 
different from that of the nearest-neighbor Ising model,
we are interested in the critical properties of 
systems with this type of interaction. 
We have previously studied phase transitions 
and the temperature dependence of ordering of model SC materials 
with specified parameters $D$ and $g$. 
In those cases, most systems exhibit a first-order phase transition, 
and the critical properties of the models were not studied in detail.   
In the present study, we investigate properties near 
the critical point in the parameter space.
In the case of the WP model, 
the critical properties are those of the short-range Ising ferromagnet. 
However, the critical properties of the present model, i.e., 
the critical exponents 
which characterize the critical universality, are expected
to be different from those of the short-range Ising model. 

The organization of the rest of this paper is as follows. 
In Sec.~\ref{sec:MM} we present the model and the computational method; 
in Sec.~\ref{sec:CP} we discuss the finite-size scaling analysis of the
critical properties; 
in Sec.~\ref{sec:SC} we discuss the spin configurations and
correlations; and in Sec.~\ref{sec:DISC} we present a summary and
discussion. A discussion of the long-range Husimi-Temperley model is
given in Appendix A, and a summary of finite-size scaling relations
for mean-field phase transitions is given in Appendix B. 

\section{Model and Method}
\label{sec:MM}

In this paper, we study the critical phenomena of models with 
elastically mediated 
spin-spin interactions on the simple square lattice (2D), and also on 
the simple cubic lattice (3D) with periodic  boundary conditions.
Here we use Monte Carlo (MC) simulations according to 
the constant-pressure method.\cite{Konishi2007} 
In the Monte Carlo simulation, 
we choose a site $i$ randomly and update the spin state $\sigma_i=(\pm 1)$ 
and the position of the molecule$(x_i,y_i,z_i)$ by the 
standard Metropolis method. 
We repeat this update $N$ times, where $N$ is the number of
lattice sites. Then, we update the volume of the total system. 
We define this sequence of procedures to be one Monte Carlo step (MCS).

Instead of the Ising-like interactions of the WP model, Eq.~(\ref{WP}),
we adopt the following elastic interactions between molecules:\cite{Konishi2007}
\begin{eqnarray}
V&=& V_{\rm nn}+V_{\rm nnn} \\
V_{\rm nn}&=& \frac{k_1}{2}\sum_{\langle i,j \rangle}[r_{ij}-(R_i+R_j)]^2 \\
V_{\rm nnn}&=&\frac{k_2}{2}
\sum_{\langle\langle i,j \rangle\rangle}[r_{ij}-\sqrt{2}(R_i+R_j)]^2,
\end{eqnarray}
where $r_{ij}$ is the distance between the $i$-th and $j$-th sites.
$V_{\rm nn}$ expresses elastic interactions between 
nearest-neighbor pairs ($\langle i,j\rangle$).
Here, $R_i$ and $R_j$ are the radii of the molecules.
The radius of each molecule is $R_{\rm HS}$ and $R_{\rm LS}$ 
for the HS and LS states, respectively. In the present work,
we set the ratio of the radii as $R_{\rm HS}/R_{\rm LS}=1.1$. 
$V_{\rm nnn}$ expresses the elastic interaction of 
next-nearest-neighbor pairs ($\langle\langle i,j\rangle\rangle$), 
which is necessary
to maintain the lattice structure but not essential for the critical behavior.
We set the ratio of the elastic constants $k_1/k_2 = 10$.
We set $k_1=40$ through out the present work.
In this study, in order to exclude other effects than those 
due to elastic interactions through 
distortion, we assume that the stiffness constants $k_1$ and $k_2$ 
do not depend on the spin state.
If we were to allow spin dependence of $k_1$ and $k_2$, 
an effective short-range interaction would appear.
In this sense, the present model treats only elastic interactions.

The order parameter for the present model is the fraction of HS molecules,
$f_{\rm HS}=N^{-1}\sum_i(2s_i-1)$. Hereafter, 
for convenience, we adopt the ``magnetization,"
\beq
M = \sum_i^Ns_i = \frac{N}{2}(f_{\rm HS} -1)
\eeq 
as the order parameter.
\begin{figure}[t]
\centerline{\includegraphics[clip,width=8cm]{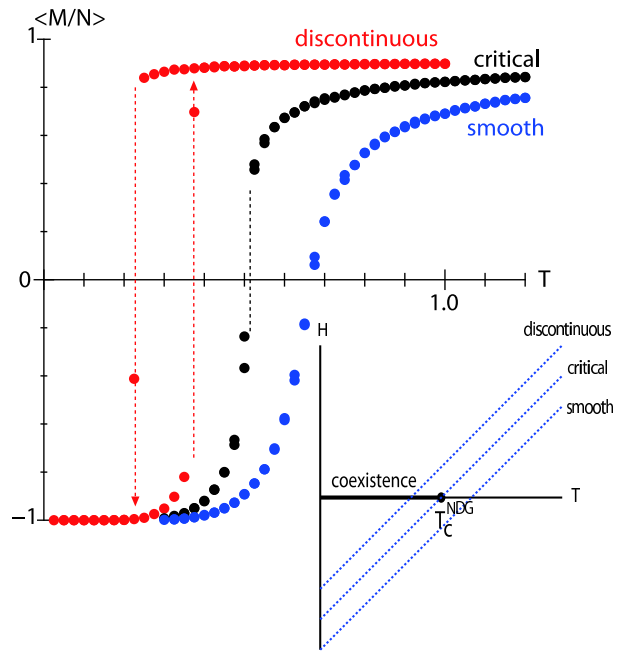} }
\caption{(Color online)
The order parameter $\langle M\rangle/N$ vs temperature $T$ 
for $D=2$, 1.51, and 1. The temperature dependences are smooth, critical, 
and discontinuous (hysteresis),
respectively. The inset shows a phase diagram of the model in the 
$(T,H)$ plane, where $T_{c}^{\rm NDG}$ is the critical 
temperature of the model ${\cal H}_{\rm elastic}$. 
The temperature changes in the model are given by the dotted lines 
in this phase diagram.
When the dotted line crosses the coexistence line denoted by the bold line, 
the system undergoes a first-order phase transition.
}
\label{D-dep}
\end{figure}
In Fig.~\ref{D-dep}, we depict the temperature dependences of 
$\langle M\rangle$ for several values of $D$. 
Here, we find the typical $D$-dependences of $\langle M(T)\rangle$.  
That is, we find a smooth dependence for large values of $D$, 
and a first-order phase transition for small $D$. 
Between them, we have a second-order phase transition.  
This $D$-dependence is understood from the phase diagram of the 
non-degenerate model (i.e. $g=1$).\cite{Nishino1}
In the present non-degenerate model 
a ferromagnetic phase transition
takes place at $T_c^{\rm NDG}$, and we expect a phase diagram 
as shown in the inset. 
In this phase diagram, $H$ is the symmetry-breaking field.

The temperature dependences of the state of the present model with 
degeneracy $g>1$ (in this work we use $g=20$) are given by 
the dotted lines in the phase diagram,
\beq
H(T)={1\over2}\left(D-k_{\rm B}T\ln g\right).
\eeq
When $D$ is larger than $D_{\rm c}=k_{\rm B}T_c\ln g$, 
the temperature dependence of $\langle M\rangle$ 
is smooth, while it shows a first-order phase transition when $D < D_{\rm c}$.
If we consider a specific material, the parameters $D$ and $g$ are given, 
and the temperature dependence of the 
state is given by one of these dotted lines. 
In most cases, the ordering is either smooth or discontinuous,
and the critical properties have therefore not yet been seriously considered. 

\section{Critical Properties}
\label{sec:CP}

We study the critical properties of the elastically 
interacting model along the coexistence line given by $T=D/\ln g$, 
i.e. $H=0$.\cite{update}   
For the WP model, the critical properties are those of the Ising model.

We next study the temperature (i.e., $T = D/k_{\rm B}T\ln g$) dependence 
of $\langle M^2\rangle$.
The spontaneous magnetization $m_s$ and the susceptibility per spin $\chi$ 
are obtained from the relation
\beq
{\langle M^2\rangle\over N^2}=m_s^2+k_{\rm B}T{\chi\over N},
\label{eq:M2}
\eeq
where $N=L^d$ is the total number of spins, and
\beq
\chi={\langle M^2\rangle\over Nk_{\rm B}T},
\eeq  
which is the susceptibility per spin above the critical point.
Here $\langle \cdots \rangle$ denotes the thermal average, i.e.,
\beq
\langle M^2\rangle={{\rm Tr}M^2e^{-\beta {\cal H}}\over 
{\rm Tr}e^{-\beta {\cal H}}}.
\eeq
\begin{figure}[t]
$$\begin{array}{c}
\includegraphics[clip,width=5.5cm]{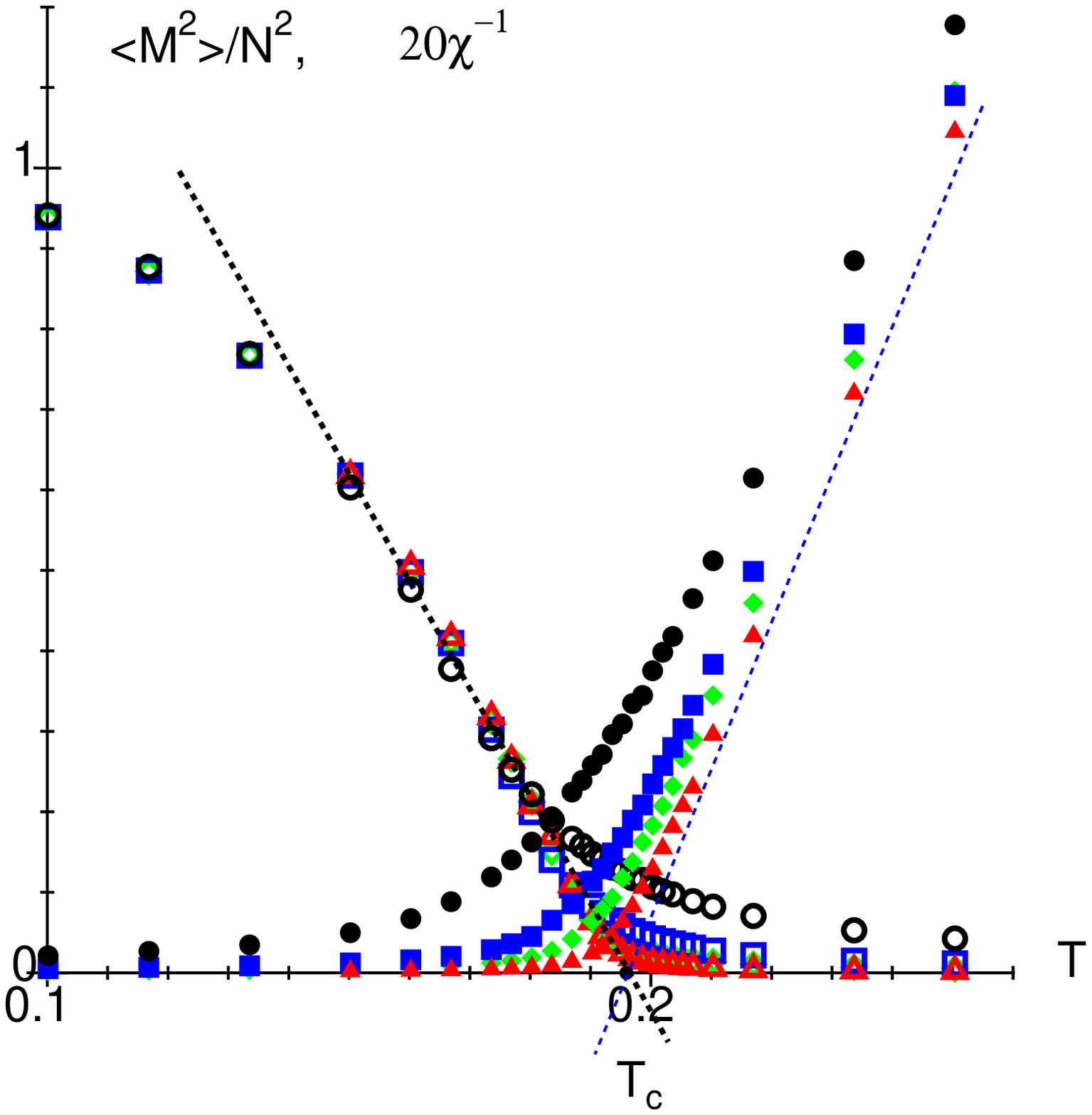} \\
({\rm a}) \\
\includegraphics[clip,width=5.5cm]{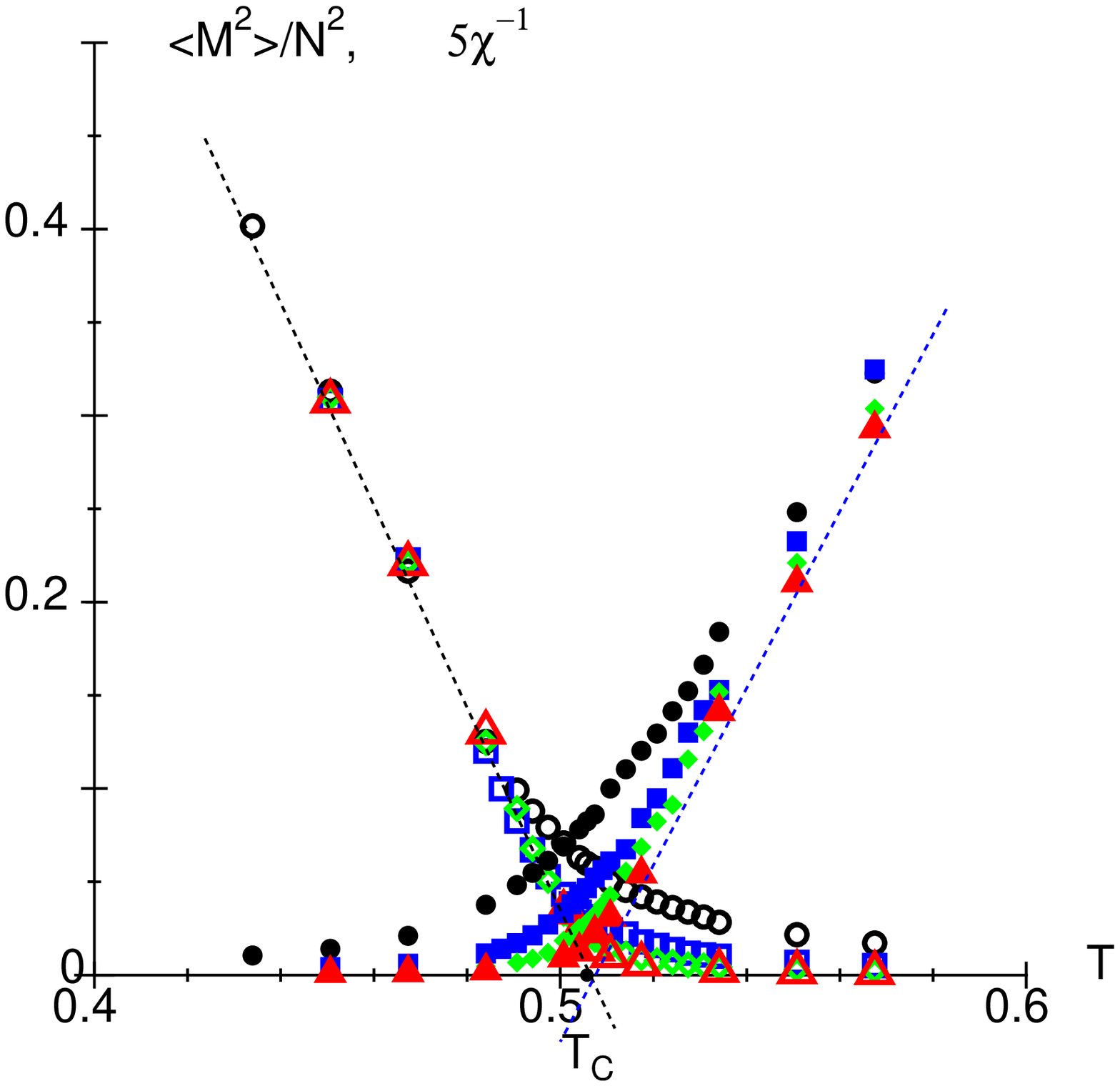} \\
({\rm b}) \end{array}$$
\caption{(Color online)
Temperature dependences of ${\langle M^2\rangle / N^2}$  and 
the inverse susceptibility $\chi^{-1} = N k_{\rm B}T/\langle M^2 \rangle$.
(a) A two-dimensional system (square lattice):  
${\langle M^2\rangle / N^2}$ for the linear sizes $L=$ 10, 20, 30, and 50 
are plotted 
by closed symbols of circle, square, diamond, and triangle, respectively; 
and $\chi^{-1}$ are plotted by the corresponding open symbols 
(multiplied by 20 for improved visibility).
(b) A three-dimensional system (simple cubic lattice):
${\langle M^2\rangle / N^2}$ for $L=$ 8, 12, 16, and 20 are plotted 
by closed symbols of circle, square, diamond, and triangle, respectively; 
and $\chi^{-1}$ are plotted by the corresponding open symbols 
(multiplied by 5 for improved visibility).
The dotted straight lines show the expected behaviors for a mean-field 
phase transition in an infinite system.
}
\label{M2T}
\end{figure}

In Fig.~\ref{M2T}, we depict the temperature dependences of 
${\langle M^2\rangle / N^2}$, and $\chi^{-1}$. 
We find a clear linear dependence of ${\langle M^2\rangle / N^2}$ 
below $T\simeq 0.20$ for the 2D model, and $T\simeq 0.51$ for the 3D model.
This linear dependence indicates that $m_{\rm s}^2\propto T_{c}-T$ and thus
the critical exponent $\beta=1/2$.  

In Fig.~\ref{M2T}, 
we also find that $\chi^{-1}$ vanishes linearly at $T_{c}$, 
which indicates $\gamma = 1$.  
This set of critical exponents agrees with those of the 
mean-field universality class.
The size dependence of the inverse susceptibility in 
Fig.~\ref{M2T} is rather large, but 
we found similar size dependences of  ${\langle M^2\rangle / N^2}$, and 
$\chi^{-1}$ 
in the long-range Husimi-Temperley model discussed in Appendix A.
This indicates that the properties shown in Fig.~\ref{M2T} are 
inherent to models in the mean-field universality 
class.\cite{Fisher1968,Brezin-ZinnJustin,Dunweg,Laradji,PRIV83,binder1985,Rikvold1993,Luijten-Bloete,Zhu,Jones} 

\subsection {Binder Plot}

\begin{figure}[t]
\centerline{\includegraphics[clip,width=5.5cm]{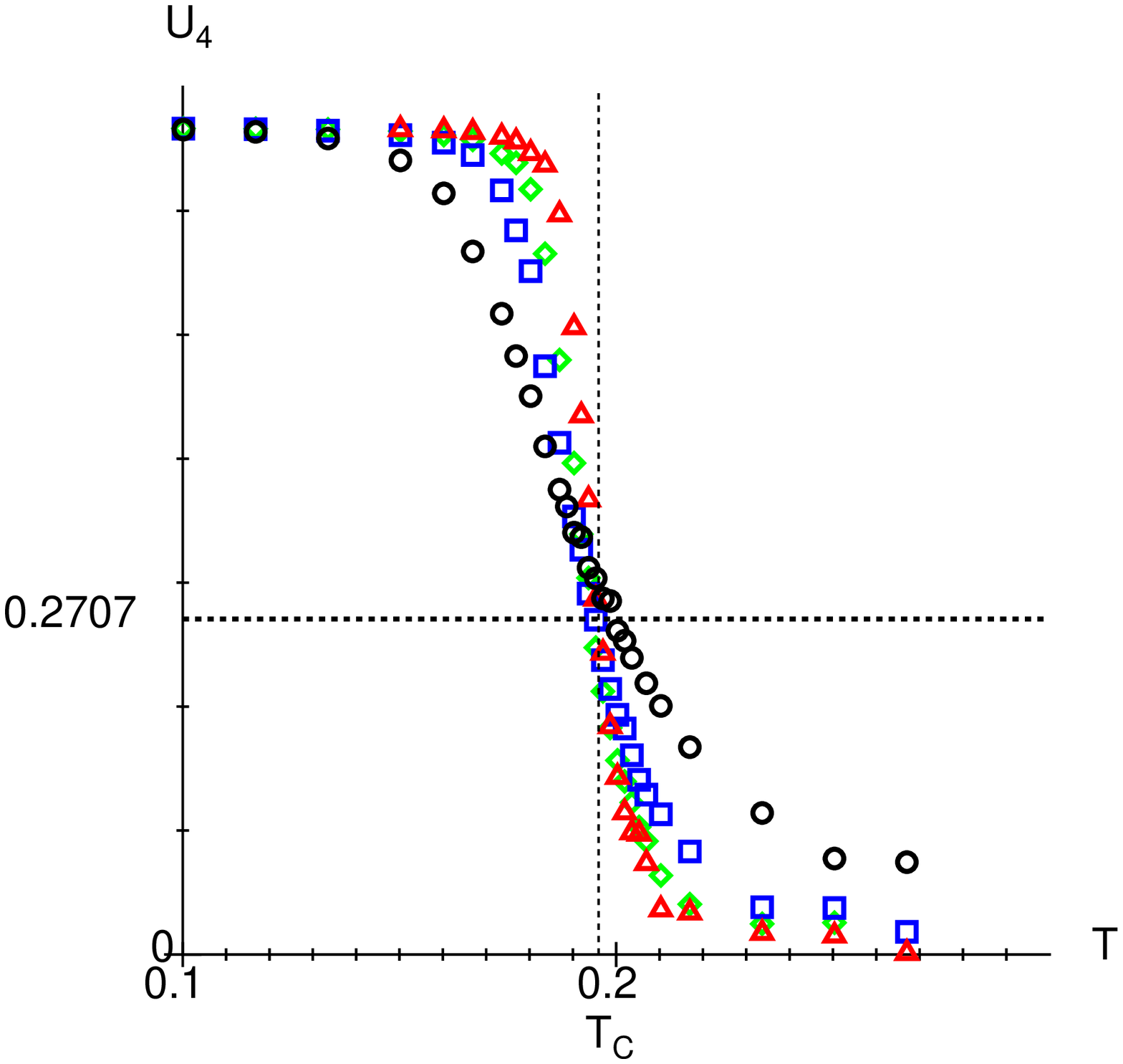} }
\centerline{(a)}
\centerline{\includegraphics[clip,width=5.5cm]{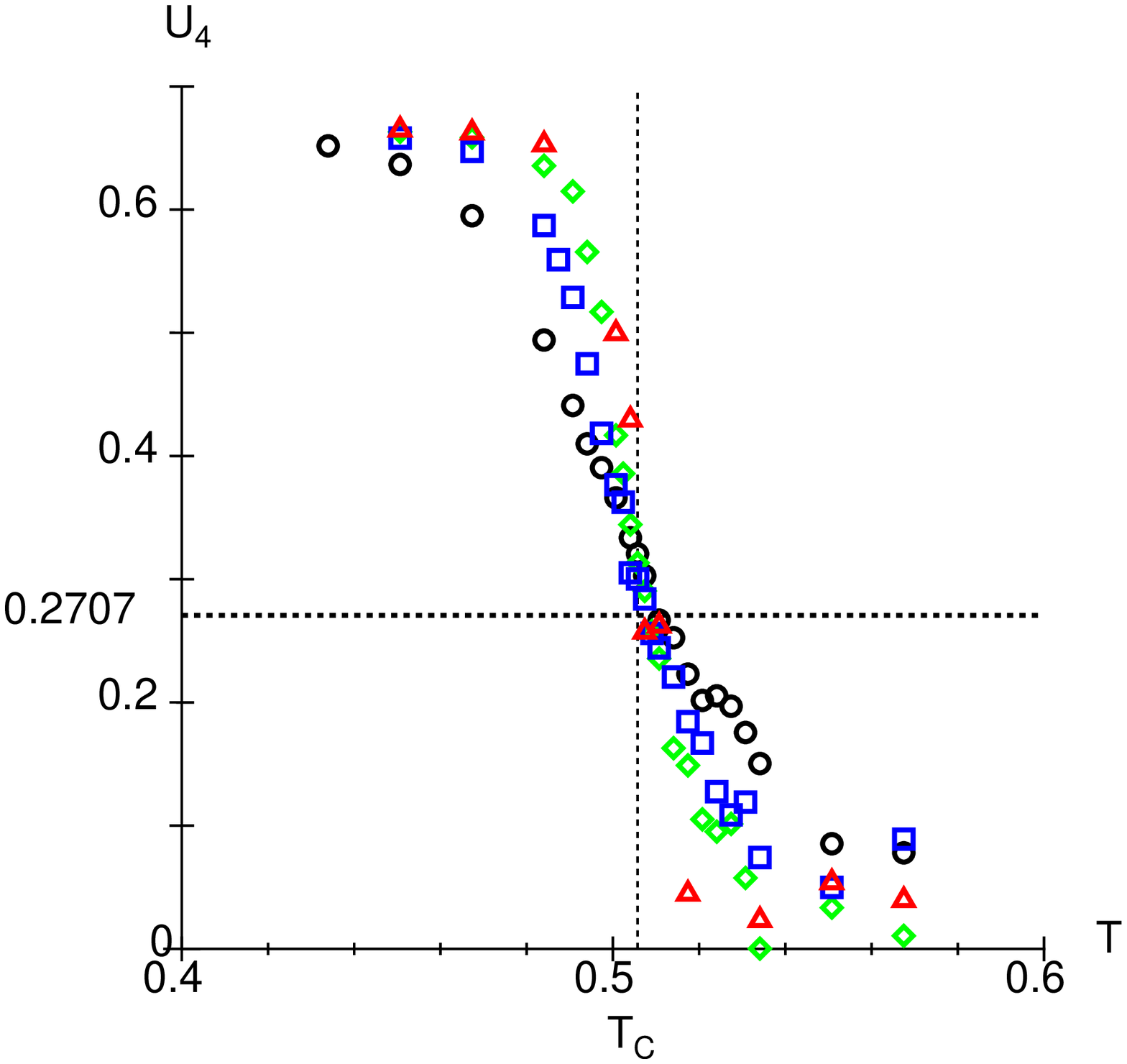} }
\centerline{(b)}
\caption{(Color online) 
Temperature dependence of the Binder cumulant.
(a) A two-dimensional system (square lattice).  
Data for $L=$ 10, 20, 30, and 50 are plotted 
as circle, square, diamond, and triangle, respectively.
(b) A three-dimensional system (simple cubic lattice).
Data for $L=$ 8, 12, 16, and 20 are plotted as 
circle, square, diamond, and triangle, respectively.
}
\label{Binder}
\end{figure}

We estimated the critical temperature by 
analysis of the Binder cumulant,\cite{binder} 
\beq
U_4=1-{\langle M^4\rangle\over 3\langle M^2\rangle^2}.
\eeq
Plotted for different system sizes, this quantity has a crossing at the 
critical point. 
It has been extensively studied for the mean-field 
universality class.\cite{Brezin-ZinnJustin}
We depict the Binder plot in Fig.~\ref{Binder}. The crossings are consistent 
with the values obtained from $\langle M^2\rangle / N^2$: 
$T_c \simeq 0.20$ for $d=2$ and $0.51$ for $d=3$.
The value of $U_4$ at the crossing is universal and independent of the 
spatial dimension. It is in excellent agreement with the theoretical
result,\cite{Brezin-ZinnJustin,Luijten1995}
\beq
U_4=1-{\Gamma^4(1/4)\over 24\pi^2}=0.27\cdots ,
\label{eq:U4star}
\eeq
where $\Gamma$ is the Gamma function. 
In Appendix A, we show the Binder plot for the long-range
Husimi-Temperley model, which gives the same fixed-point value. 

\subsection{Finite-size scaling}


Finite-size scaling is one of the most useful methods to extract
critical properties for infinite systems from numerical data for
finite systems.\cite{PRIV84,PRIV90} However, special caution
must be used when considering transitions in the mean-field
universality class, which do not obey the hyperscaling relation 
$2 \beta + \gamma = d \nu$ 
that relates the critical correlation-length exponent $\nu$ 
with the spatial dimensionality $d$ for transitions with nonclassical
exponents.\cite{GOLD92} Essentially, lengths are not well defined
in systems with mean-field phase transitions, and the linear system size
$L$ is replaced by the number of sites $N$ as the fundamental
finite-size scaling variable. A particularly clear example is the long-range
Husimi-Temperley model discussed in Appendix A, 
in which every spin interacts with every other
with a strength proportional to $1/N$. The finite-size scaling variable that
replaces the standard $t L^{1/\nu}$ is $t N^{1/2}$.\cite{PRIV83,binder1985}
This corresponds to
an {\it effective\/} exponent $\nu^* = 2/d$,\cite{PRIV83,Luijten-Bloete} 
different from the value of $\nu= 1/2$, obtained from the Gaussian 
approximation.\cite{GOLD92} 
An effective exponent for the correlation function on the large scales that
are relevant for finite-size scaling is obtained from $\nu^*$ by
the standard exponent relation $\eta^* = 2 - \gamma / \nu^* = (4-d)/2$. 
Thus one expects the scaling expression 
$\langle M^2 \rangle 
= L^{d + 2 - \eta^*} \mathcal{M}^2(t L^{1/\nu^*}) 
=L^{3d/2} \mathcal{M}^2(tL^{d/2})$, where 
$\mathcal{M}^2$ is a scaling function. 
A summary of the mechanisms that lead to these results is given in Appendix B.

In Fig.~\ref{Fig_Binder-scale3d} we demonstrate that 
the Binder cumulants for different $L$ collapse onto
a single scaling function when plotted vs $tL^{d/2}$, 
\begin{figure}[t]
\centerline{\includegraphics[clip,width=5.5cm]{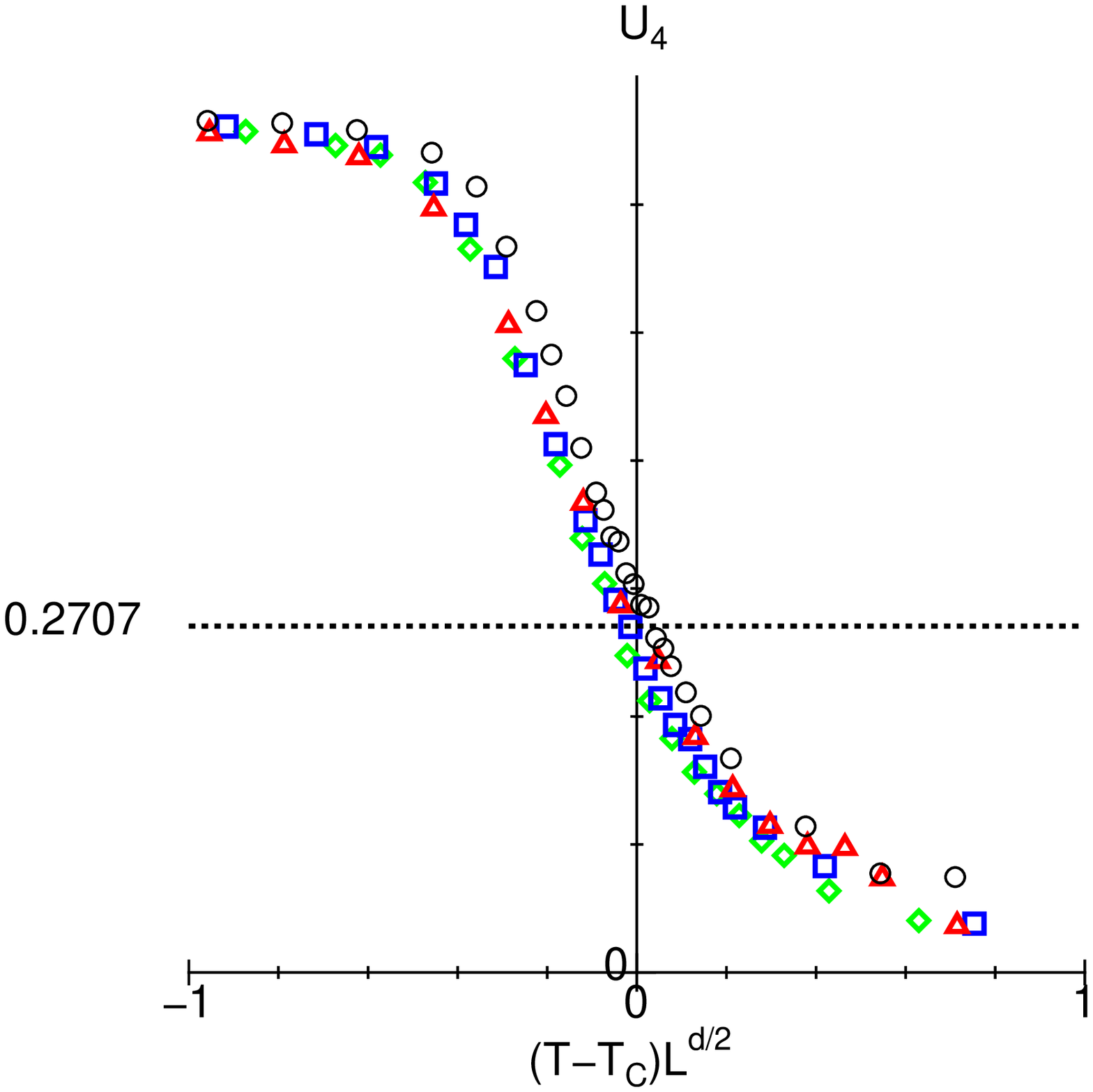} }
\centerline{(a)}
\centerline{\includegraphics[clip,width=5.5cm]{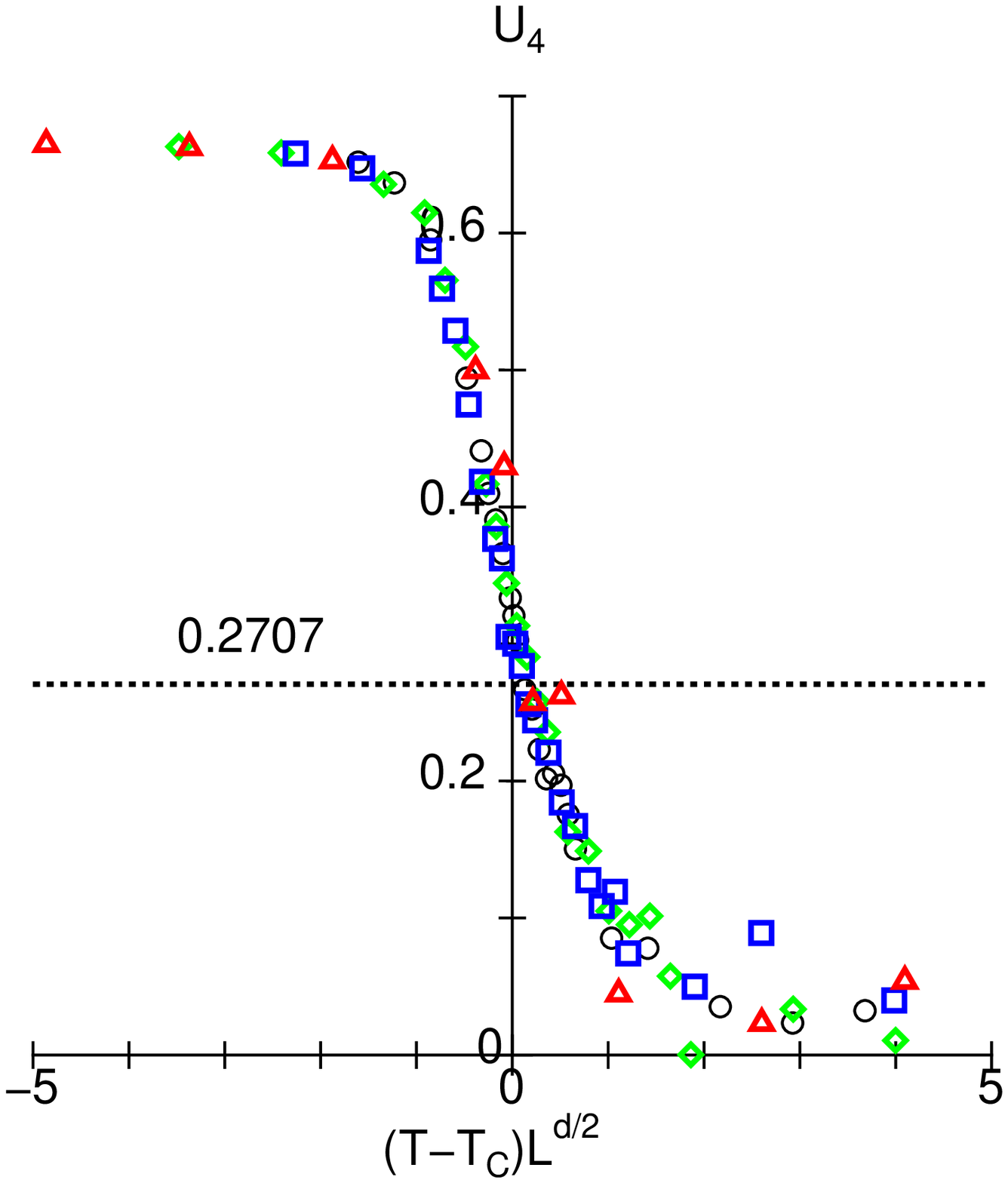} }
\centerline{(b)}
\caption{(Color online)
Finite-size scaling plots of the Binder cumulant $U_4$ vs $(T-T_c)L^{d/2}$. 
(a) The two-dimensional system.
Data for $L=$ 10, 20, 30, and 50 are plotted 
as circle, square, diamond, and triangle, respectively.
(b) The three-dimensional system. 
Data for $L=$ 8, 12, 16, and 20 are plotted as 
circle, square, diamond, and triangle, respectively.
}
\label{Fig_Binder-scale3d}
\end{figure}
and in Fig.~\ref{Fig_scaleM2} we plot the finite-size scaling functions 
for $\langle M^2 \rangle$.
In both cases we find good data collapse, both for $d=2$ and $d=3$. 
These finite-size scaling relationships are also 
seen in the long-range Husimi-Temperley model discussed in Appendix A. 
\begin{figure}[t]
\centerline{\includegraphics[clip,width=5.5cm]{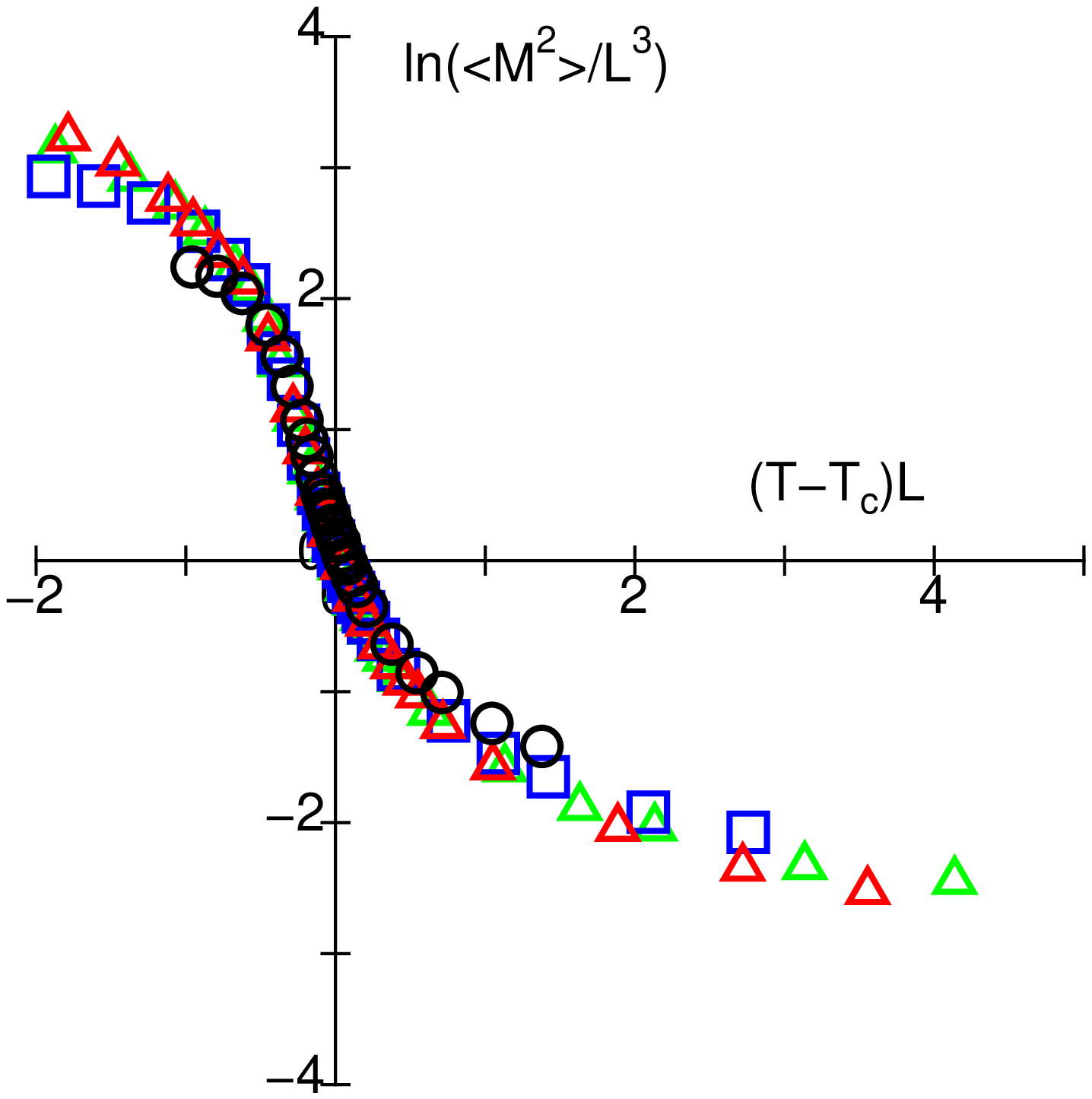} }
\centerline{(a)}
\centerline{\includegraphics[clip,width=5.5cm]{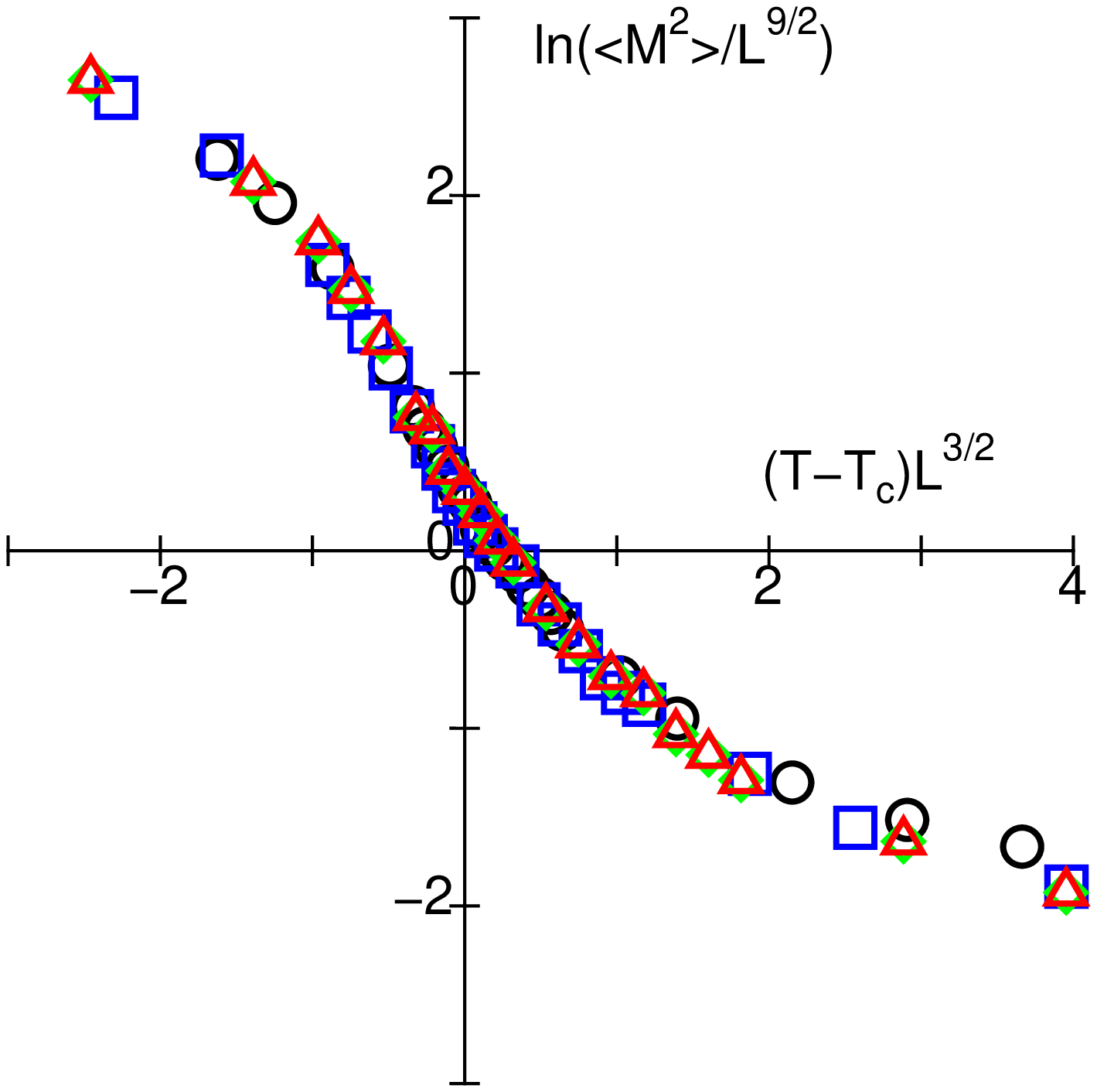} }
\centerline{(b)}
\caption{(Color online)
Finite-size scaling plots of $\langle M^2 \rangle$.
(a) A two-dimensional system (square lattice).  
Data for $L=$ 10, 20, 30, and 50 are plotted 
as circle, square, diamond, and triangle, respectively.
(b) A three-dimensional system (simple cubic lattice).
Data for $L=$ 8, 12, 16, and 20 are plotted as 
circle, square, diamond, and triangle, respectively.
}
\label{Fig_scaleM2}
\end{figure}

\subsection{Phenomenological scaling analysis}

\begin{figure}[t]
\centerline{\includegraphics[clip,width=8.0cm]{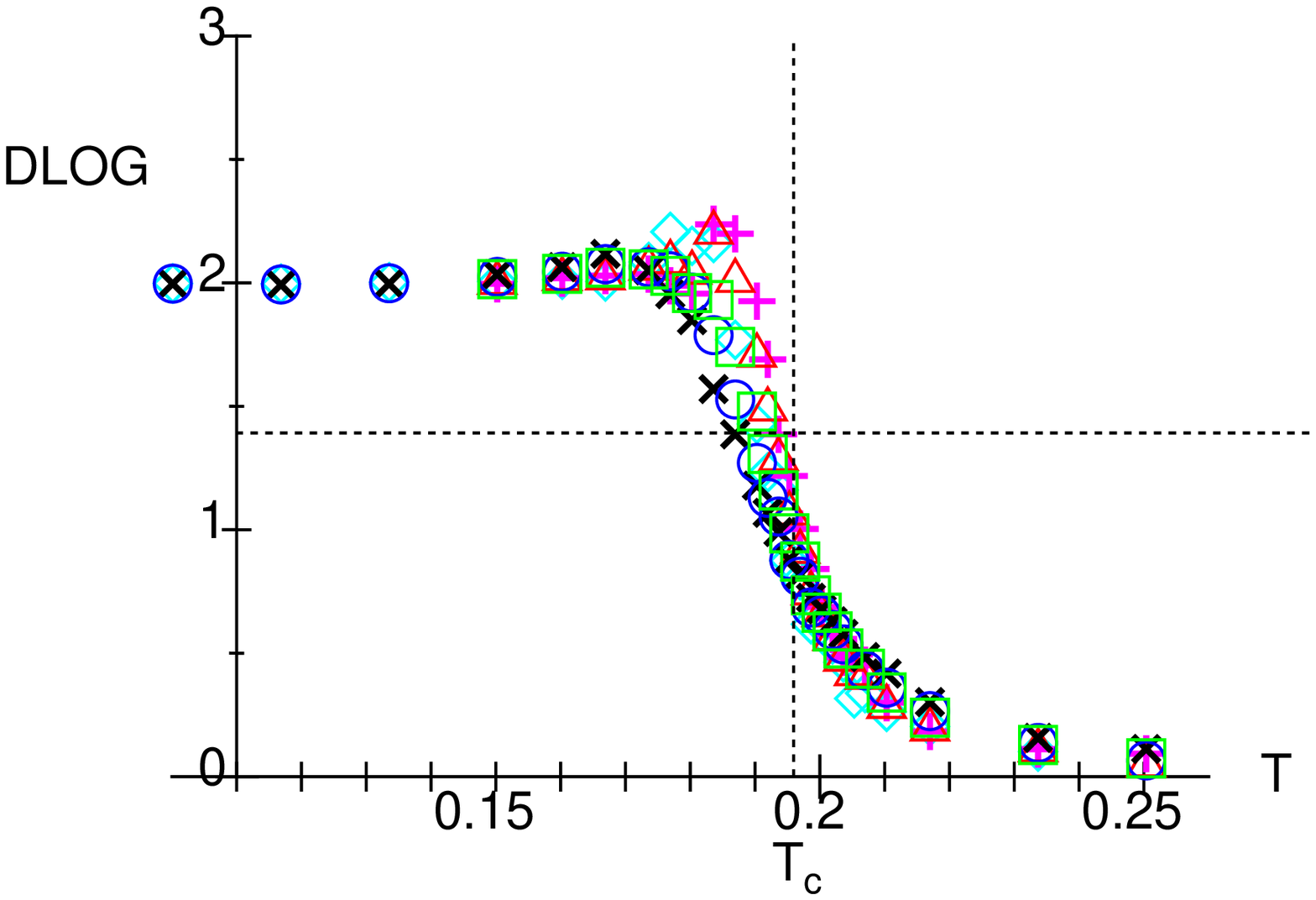} }
\centerline{(a)}
\centerline{\includegraphics[clip,width=8.0cm]{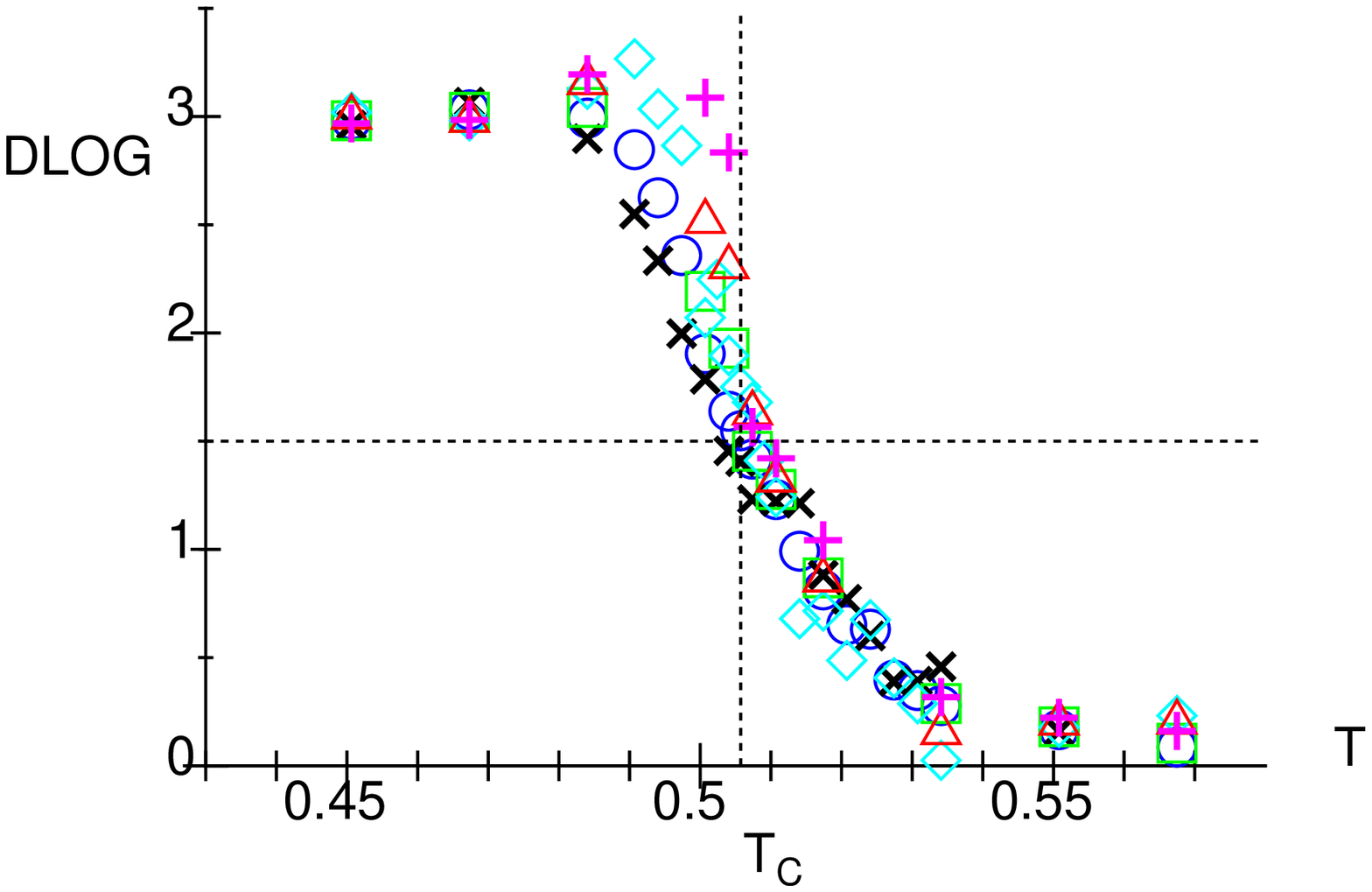} }
\centerline{(b)}
\caption{(Color online) 
Phenomenological Monte Carlo renormalization plots of $\langle
M^2\rangle/N$, as defined in Eq.~(\protect\ref{eq:DLOG}).
(a) A two-dimensional system (square lattice).  
Data for $(L,L')=$ (10,20), (10,30), (10,50), (20,30), (20,50), and (30,50)
are plotted as cross, circle, square, diamond, triangle, and plus, respectively.
(b) A three-dimensional system (simple cubic lattice).
Data for $(L,L')=$ (8,12), (8,16), (8,20), (12,16), (12,20), and (16,20)
are plotted as cross, circle, square, diamond, triangle, and plus, respectively.
}
\label{DLOG}
\end{figure}

In order to determine the critical temperature and the exponent 
$\eta^* = 2 - \gamma/\nu^*$, 
the so-called phenomenological Monte Carlo renormalization plot 
is often useful.\cite{PMCRG} 
That is, we plot
\beq
{\rm DLOG}={\ln\left(\langle M^2\rangle_L/\langle M^2\rangle_{L'}\right)\over\ln(L/L')}
-d
\label{eq:DLOG}
\eeq
as a function of $T$. 
The data for different sets of $L$ and $L'$ 
are expected to cross at a point which gives 
$T_{c}$ and $\gamma/\nu^* = 2-\eta^*$. 
In Fig.~\ref{DLOG}, we plot the temperature dependence of this 
quantity for two- and three-dimensional systems. We find a crossing in each 
figure at the position estimated by the values obtained in previous subsections: 
in the two-dimensional case, 
\beq
T_c\simeq 0.20\quad {\rm and} \quad \eta^* \simeq 1,
\eeq
and 
in the three-dimensional case,
\beq
T_c\simeq 0.51\quad {\rm and} \quad \eta^* \simeq 0.5.
\eeq
We find a similar dependence in the Husimi-Temperley model given in
Appendix A. 

\section{Spin Configuration}
\label{sec:SC}

\subsection{Spin correlation function}

\begin{figure}[t]
\centerline{ (a) \includegraphics[clip,width=5.5cm]{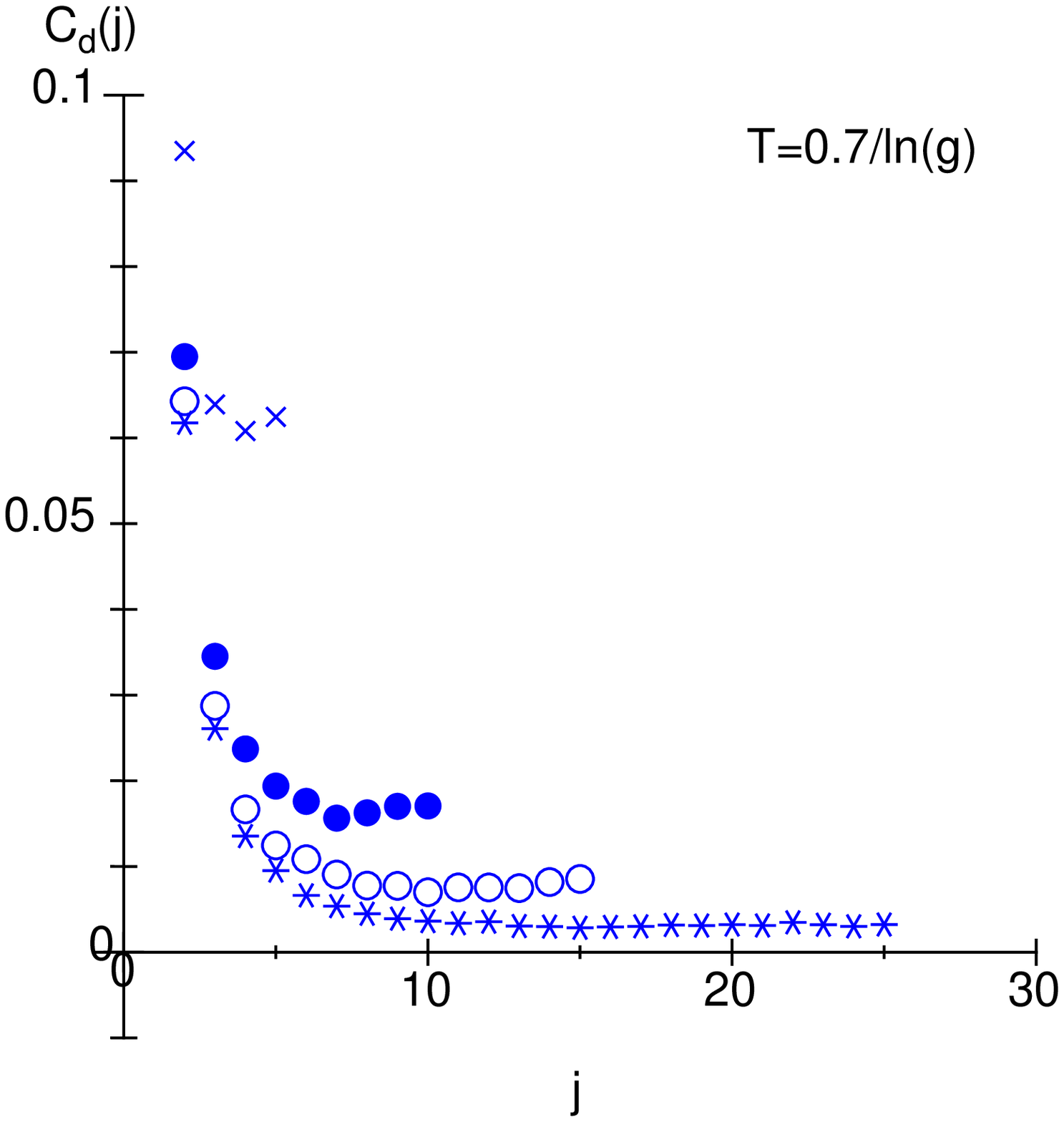} }
\centerline{ (b) \includegraphics[clip,width=5.5cm]{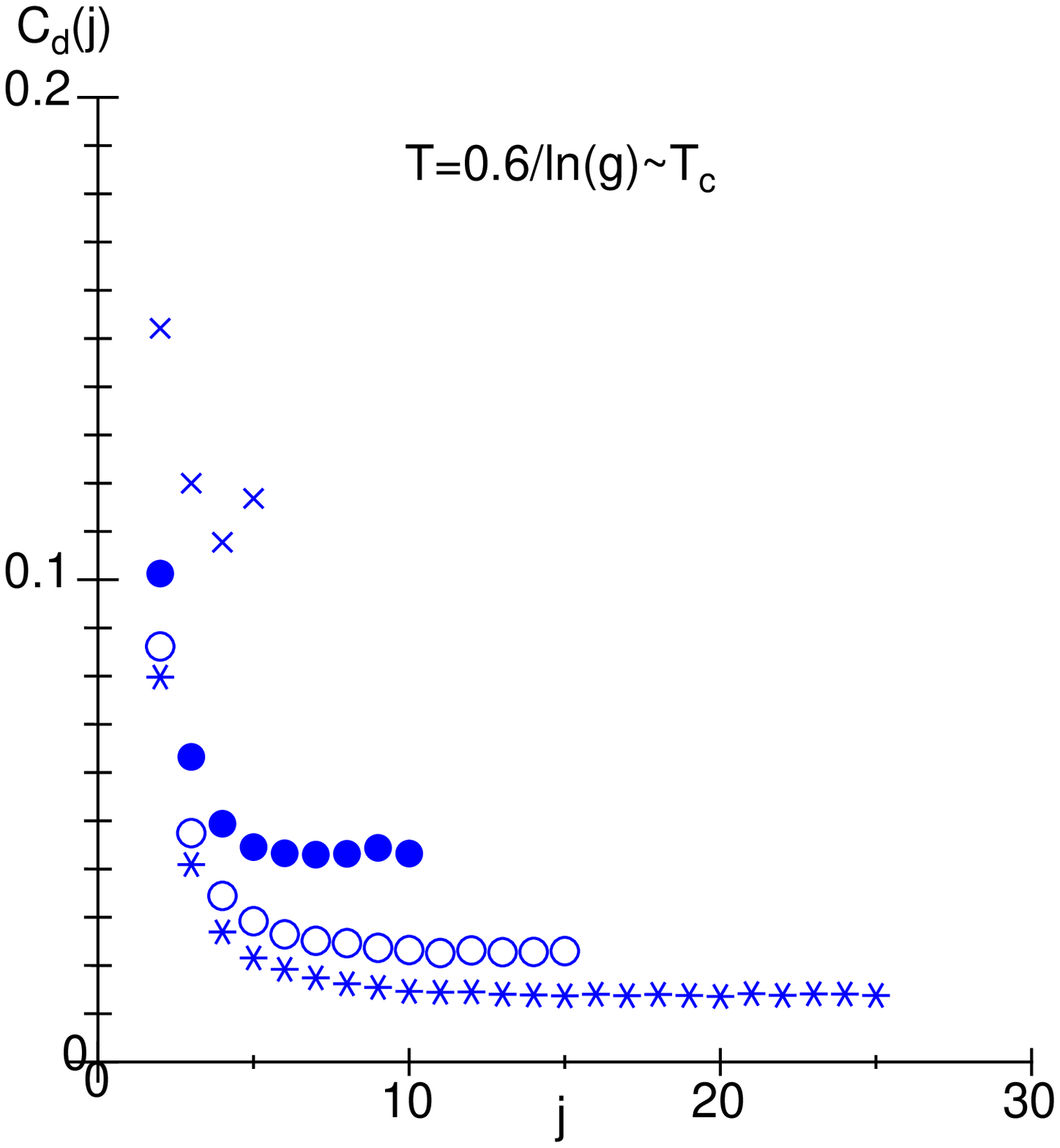} }
\centerline{ (c) \includegraphics[clip,width=5.5cm]{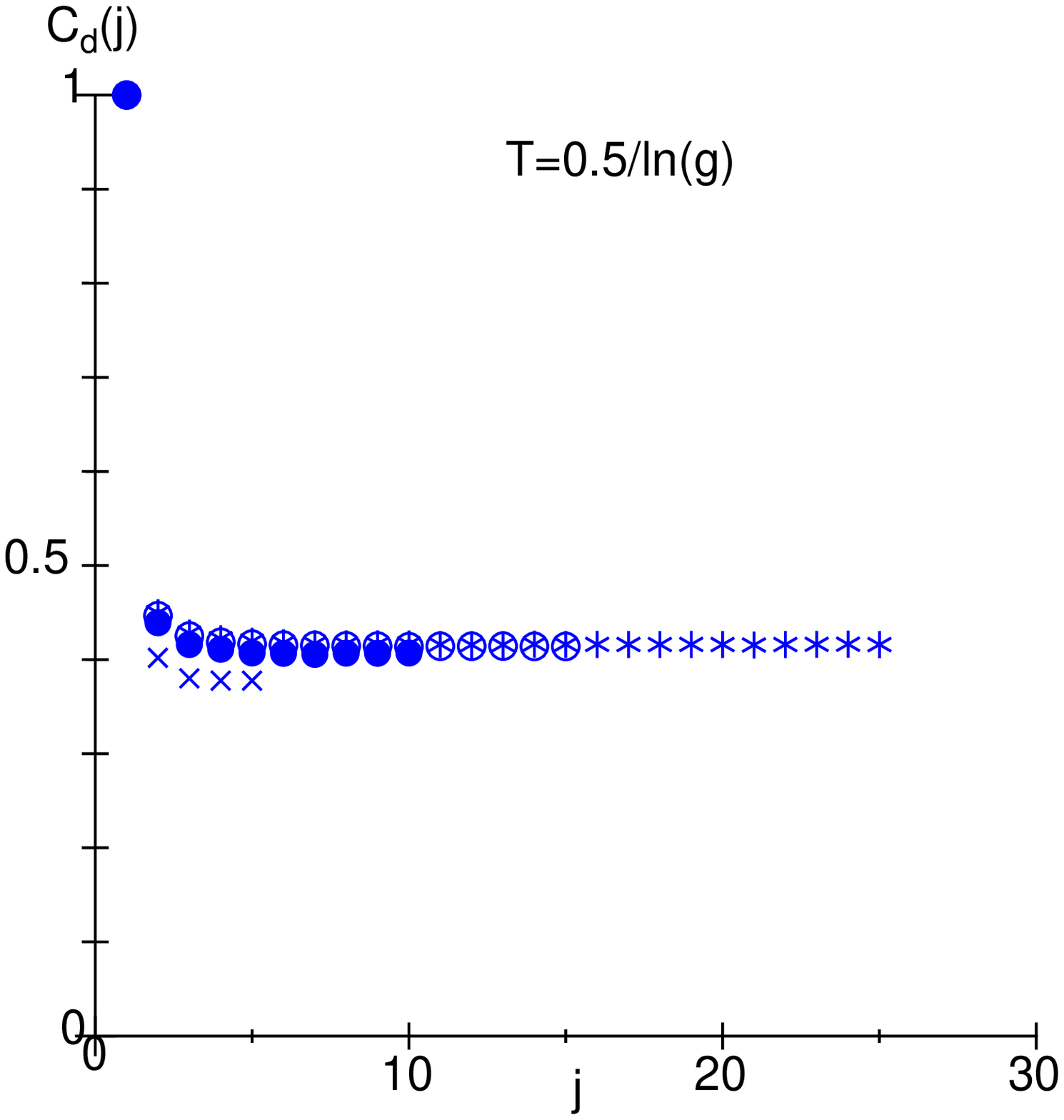} }
\caption{
(Color online)
The size and distance dependences of the 
spin correlation function in the two-dimensional model at several temperatures.
(a) $T=0.7/\ln(g)$, which is in the paramagnetic phase.
(b) $T=0.6/\ln(g)\sim T_{c}$.
(c) $T=0.5/\ln(g) < T_{c}$ in the ordered phase.
In these figures, the system sizes are $L=10~(\times)$, 
$20~(\bullet)$, $30~(\circ)$, and $50~(\ast)$.
}
\label{cor-T}
\end{figure}

Here, we concentrate on the characteristics of the spin correlation function. 
In Fig.~\ref{cor-T}, we depict the size and distance 
dependences of the correlation functions for various values of $T$: 
$T=0.7/\ln(g)$, which is in the paramagnetic phase,
$T=0.6/\ln(g) \approx T_{c}$, and  $T=0.5/\ln(g) < T_{c}$.
We plot the correlation function along the diagonal direction, i.e.,
$C_{\rm d}(r)=\langle \sigma(x_0,y_0)\sigma(x_0+r,y_0+r)\rangle$.\cite{shotrange-order}
We find unusual spin correlation functions in the disordered phase.
In short-range interaction models the correlation function decays exponentially.
In contrast, we here find the correlation to be nonzero and almost constant  
at long distances in the disordered phase at $T=0.7/\ln(g)$.
This observation indicates that the spins are strongly correlated, 
even at high temperatures. 
In the disordered phase, the susceptibility is an extensive quantity, 
and thus the total sum of the spin 
correlation function must be proportional to $N$:
\beq
N\chi T=\sum_{i}^N\sum_j^N\langle\sigma_i\sigma_j\rangle \propto N.
\eeq 
In order to satisfy this property, the constant value of the 
correlation function at long distances, $c_0$,  must depend on the 
system size as 
\beq
c_0\propto {1\over N}.
\label{corpara}
\eeq
This is in stark contrast to the result for Ising models with short-range 
interactions, $c_0 \sim e^{-L/2\xi}$ with a correlation length $\xi$ of order
unity. 

At the critical point ($T=0.6/\ln(g)\sim T_{c}$), 
the size dependence of $c_0$ is given by
\beq
c_0\propto {1\over \sqrt{N}}.
\label{cortc}
\eeq
This constant component at the critical point was pointed out by 
Luijten and Bl\"ote.\cite{Luijten-Bloete}
These observations are qualitatively different from those of the short-range
Ising model.
In the ordered state ($T=0.5/\ln(g)$), $c_0$ is independent of $N$, which
corresponds to spontaneous order.

\subsection{Spin configuration in equilibrium}

\begin{figure}[t]
$$\begin{array}{cc}
\includegraphics[width=4.5cm]{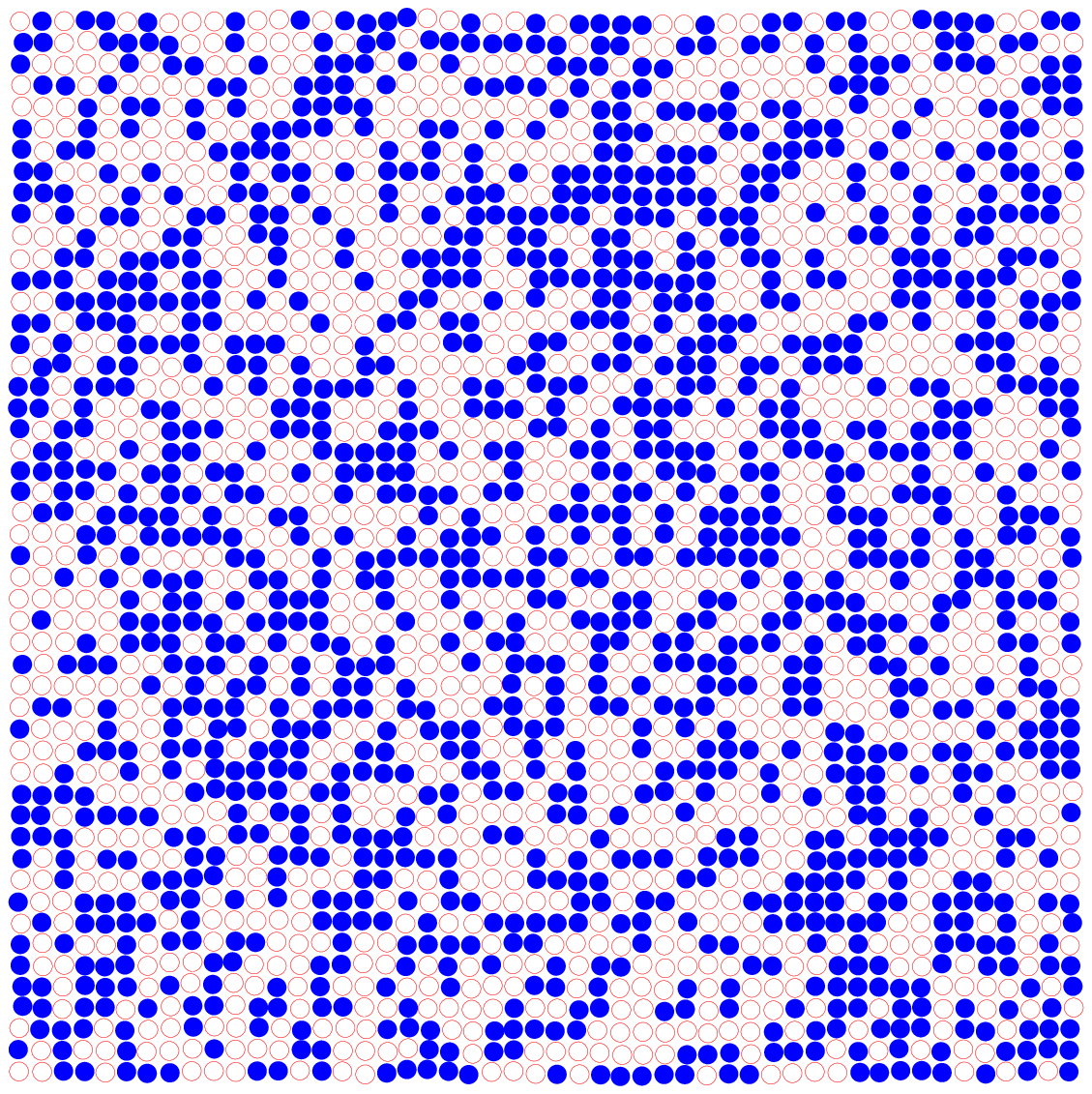} &
\includegraphics[width=4.5cm]{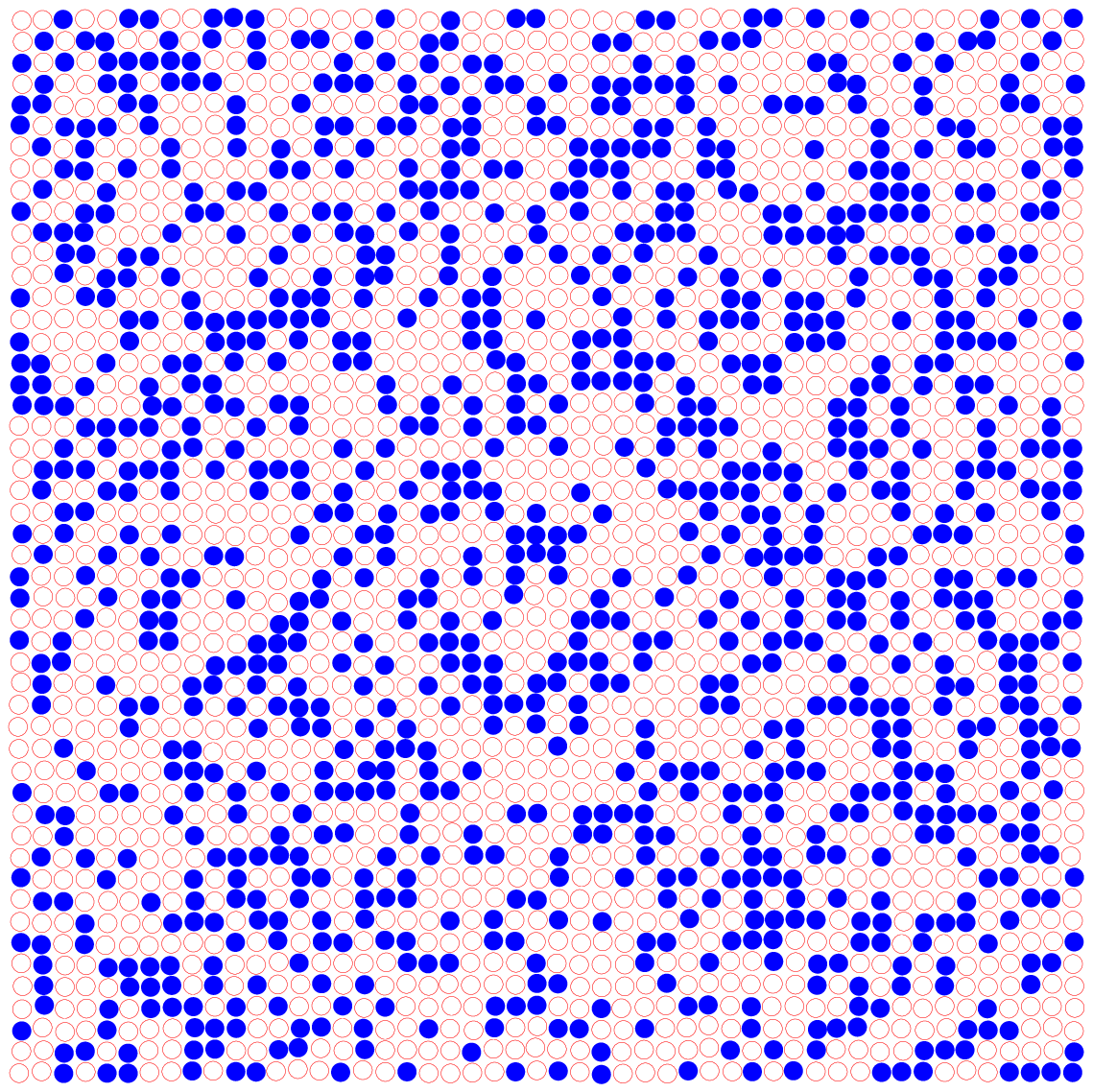} \\
({\rm a}) & ({\rm b})\\ &\\
\includegraphics[clip,width=4.5cm]{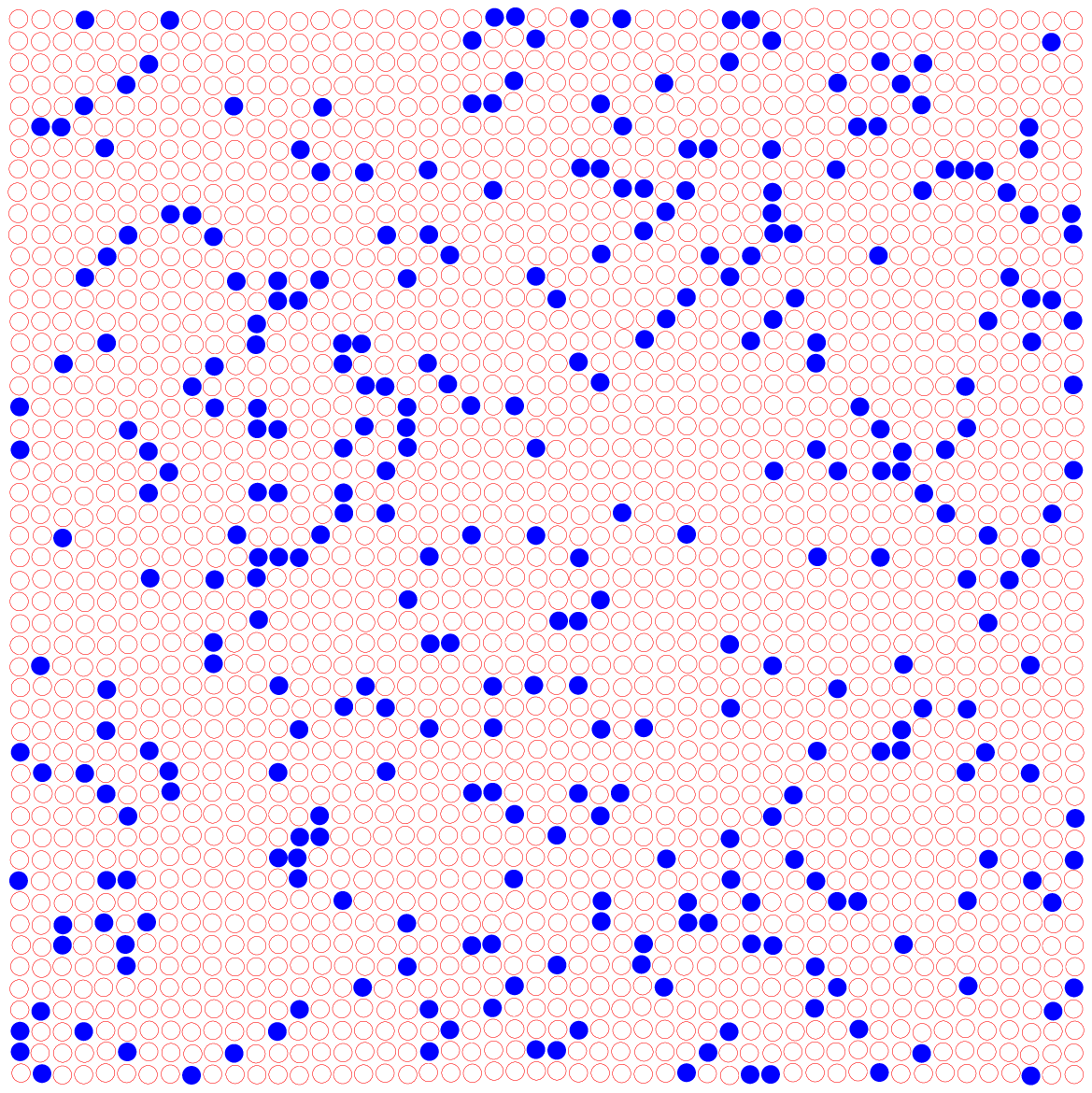} &
\includegraphics[width=4.5cm]{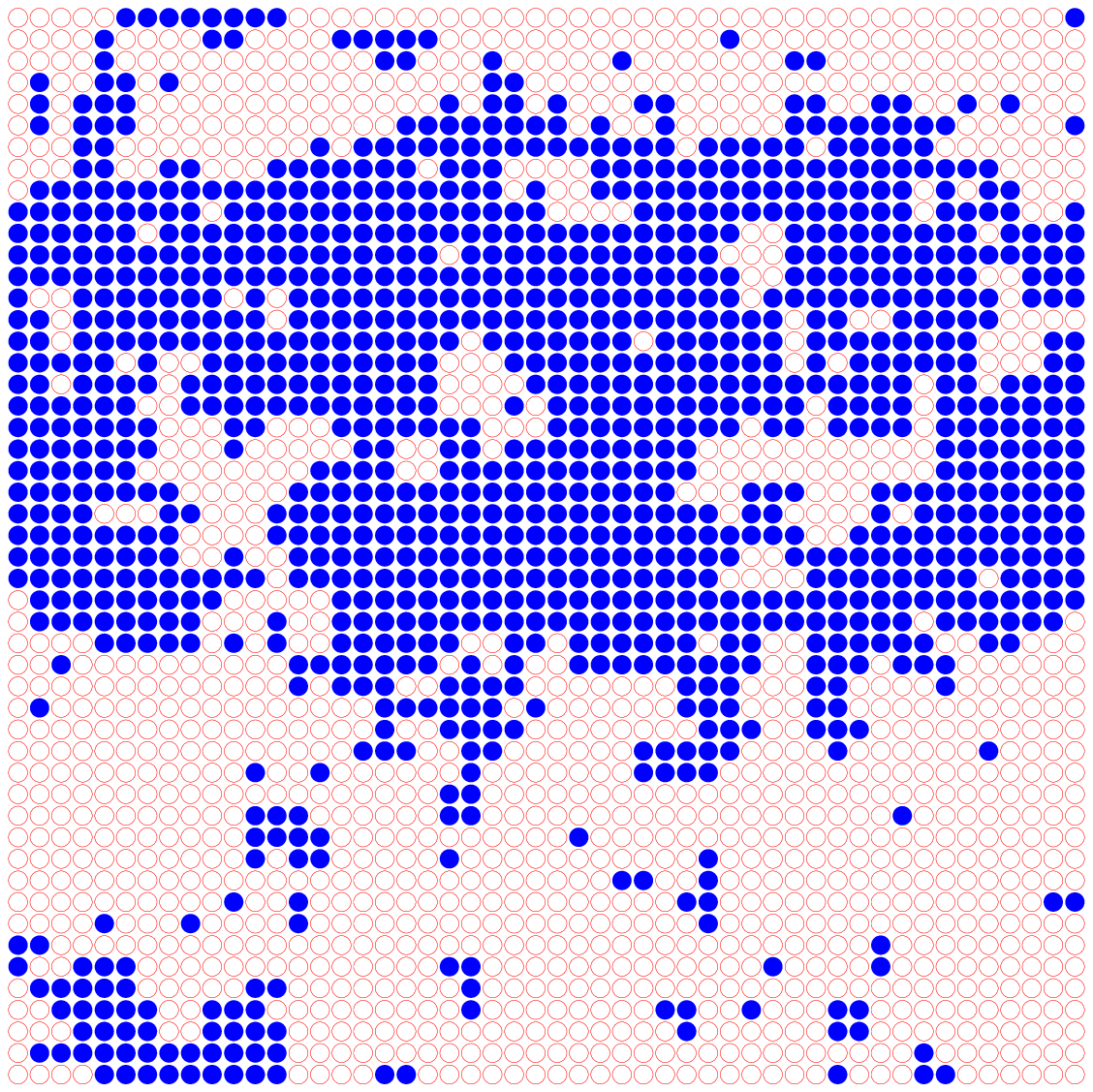} \\
({\rm c}) & ({\rm d})\end{array}$$
\caption{
(Color online)
Snapshots of equilibrium configurations of the two-dimensional model at
(a) $T=1.2T_{c}$, (b) $T=T_{c}$, and (c) $T=0.8T_{c}$. 
(d) A snapshot of the nearest-neighbor Ising ferromagnet near
the critical point, $T=2.3J$.}
\label{CONF}
\end{figure}

Next, let us discuss the characteristics of the spin configurations 
in the model.
In Fig.~\ref{CONF},
we depict three snapshots of spin configurations (a) at a high temperature, 
(b) near the critical point, and (c) at a low temperature.
We find that there are no large domain structures, even near the critical point.
For comparison, we depict a configuration at the 
critical point of the two-dimensional nearest-neighbor Ising model (d). 
The difference is striking. We also found the structure factor to be 
almost wave-number independent (not shown).  From these observations, 
we expect that usual critical behavior associated with two-phase coexistence 
will be suppressed in the present model.

\subsection{Spin configuration at the end of the hysteresis loop}

\begin{figure}[t]
\centerline{\includegraphics[clip,width=8cm]{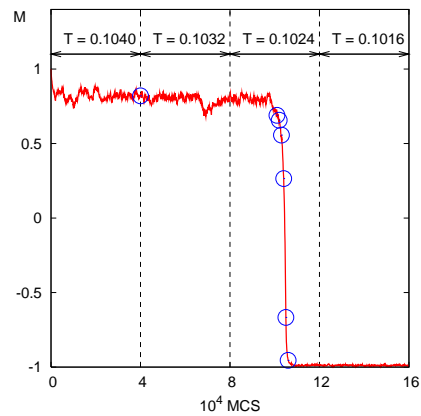} }
\caption{
(Color online)
Time dependence of the magnetization at the end of a hysteresis loop.}
\label{hyst-m}
\end{figure}

\begin{figure}[t]
$$\begin{array}{cc}
\includegraphics[width=4.5cm]{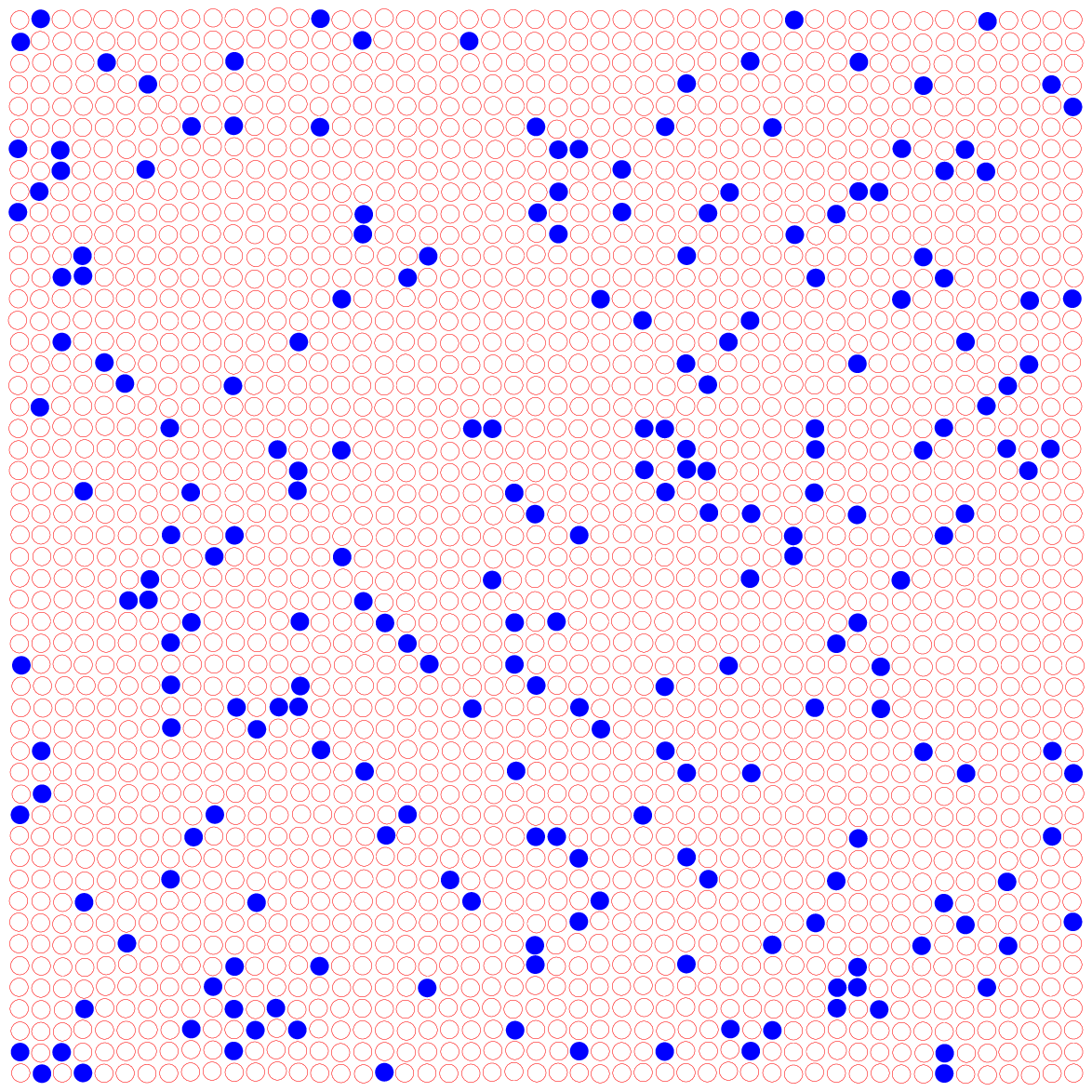} &
\includegraphics[width=4.5cm]{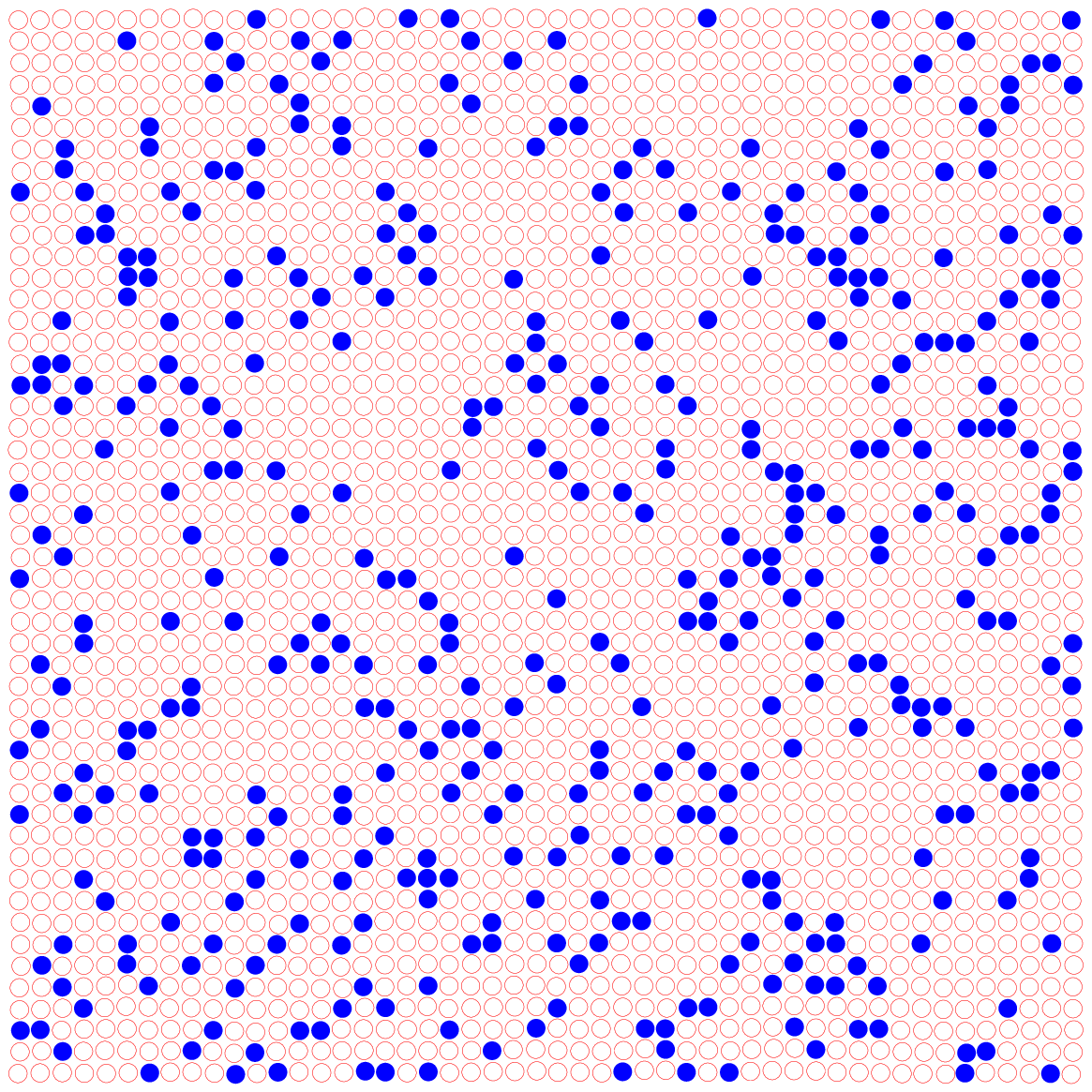} \\
({\rm a}) & ({\rm b})\\ &\\
\includegraphics[width=4.5cm]{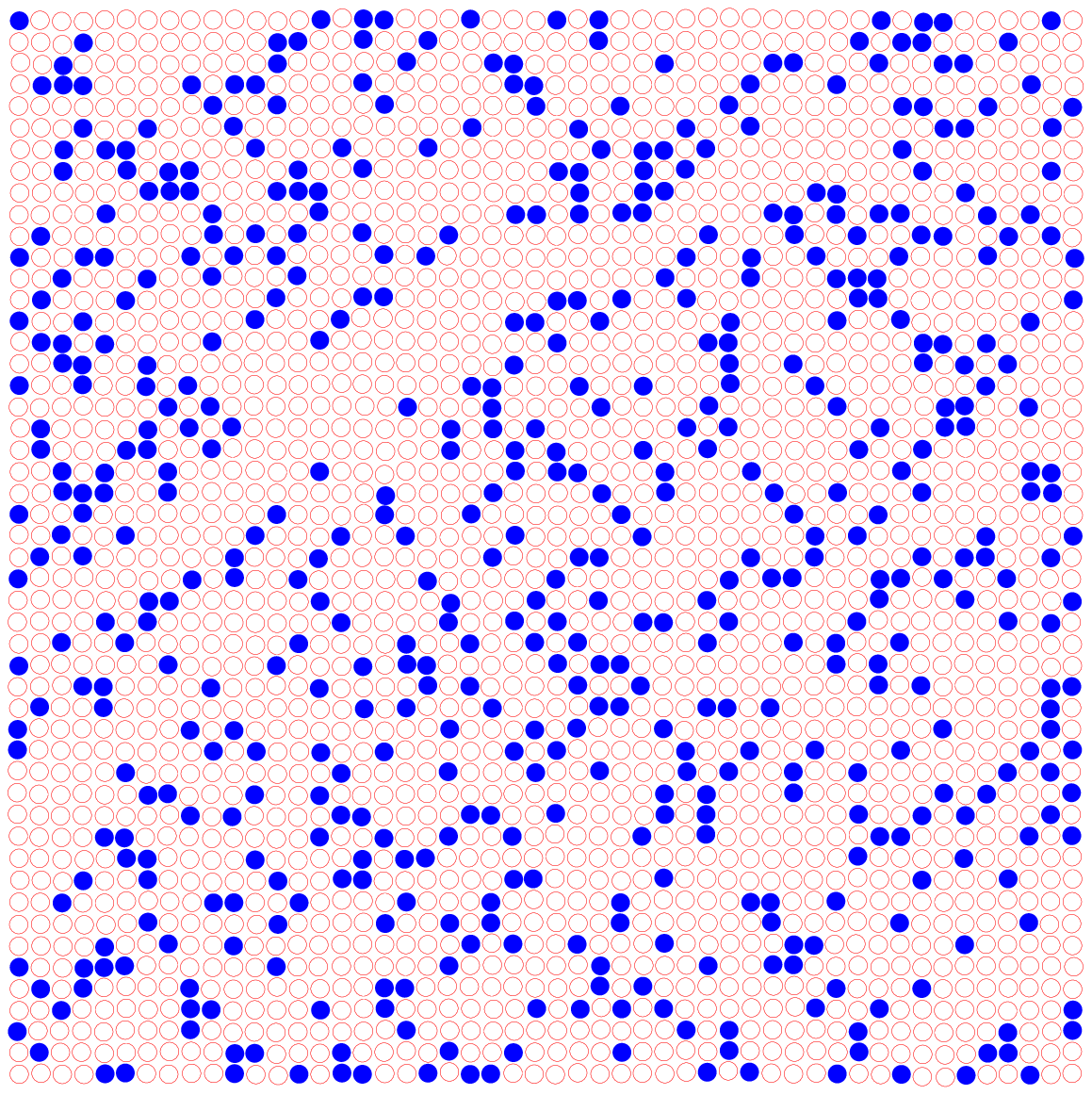} &
\includegraphics[width=4.5cm]{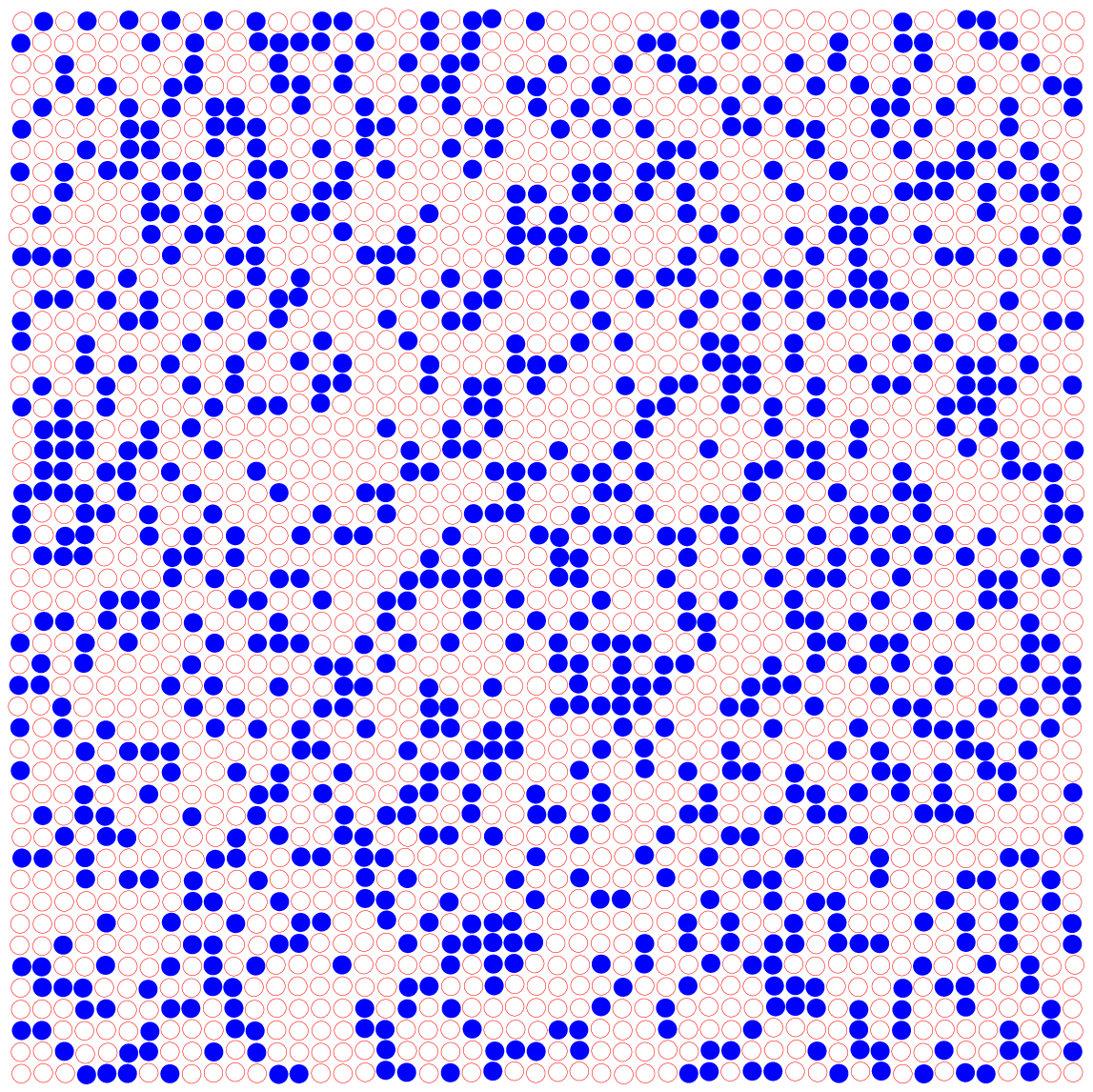} \\
({\rm c}) & ({\rm d})\\ &\\
\includegraphics[width=4.5cm]{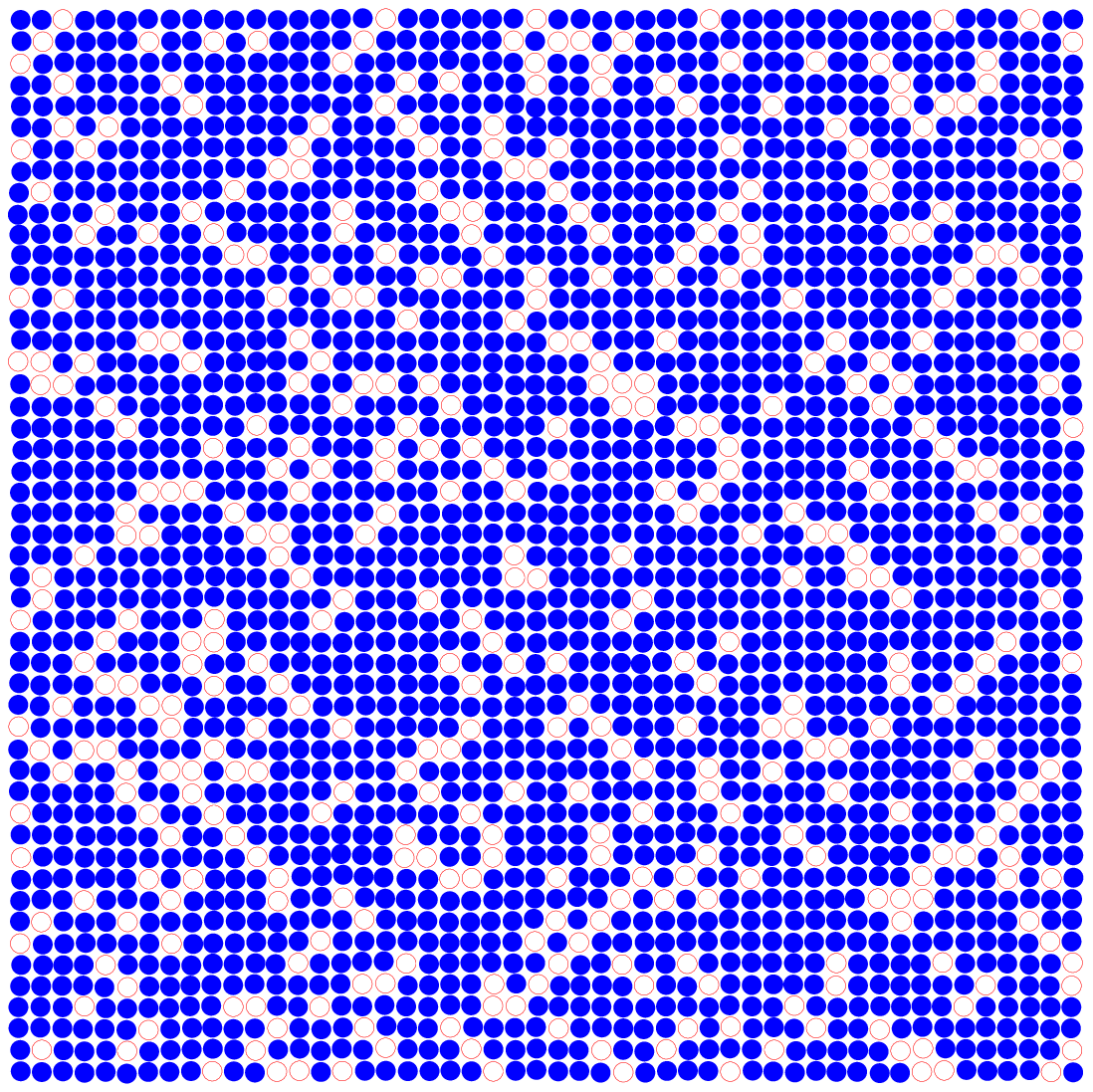} &
\includegraphics[width=4.5cm]{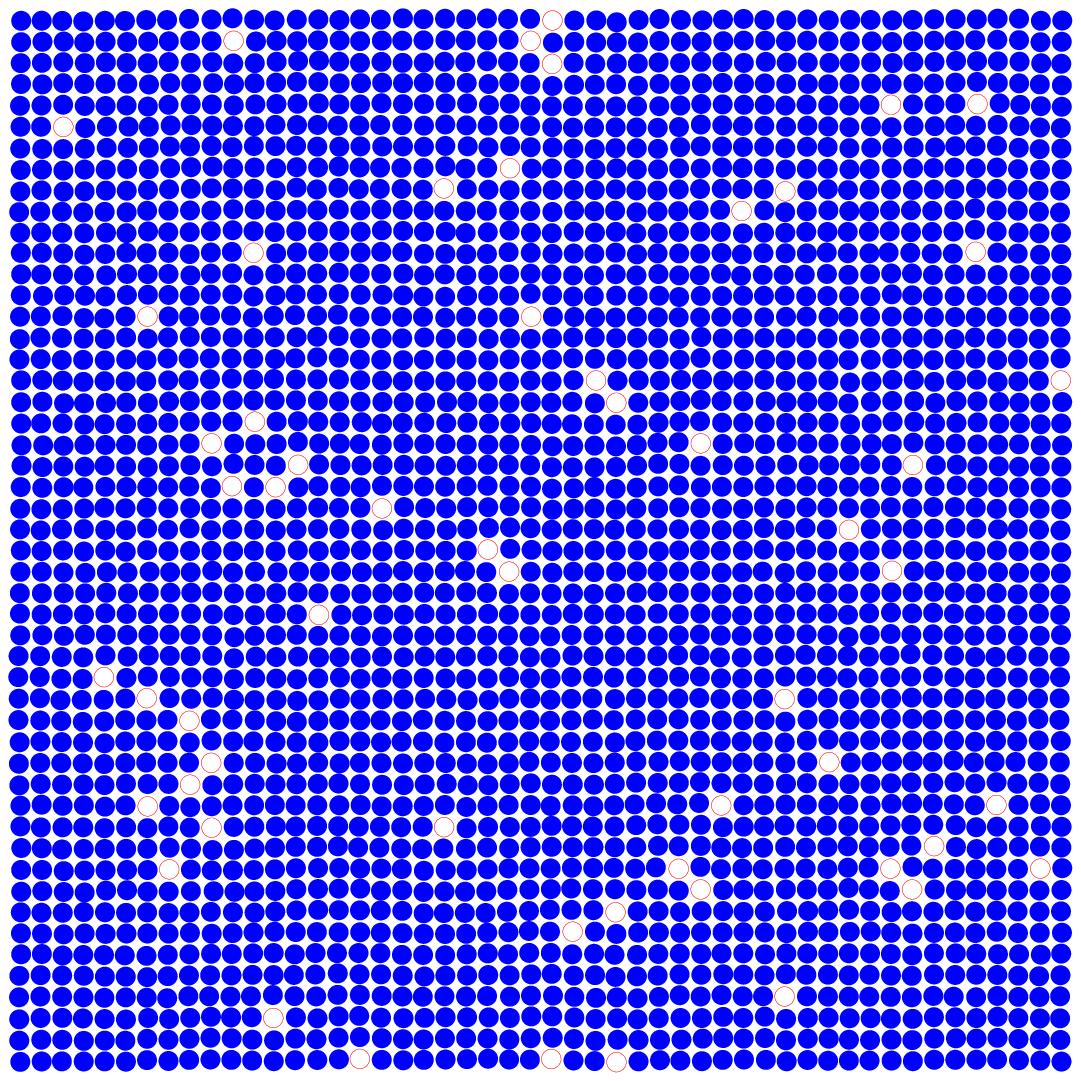} \\
({\rm e}) & ({\rm f})\end{array}$$
\caption{
(Color online)
Time dependence of the spin configuration at the end of a hysteresis loop. 
(a) Configuration just before the end point, at $T=0.1040$. 
Here no change occurs after 40000~MCS. 
(b)At $T=0.1024$. At this temperature, a sudden change of the magnetization
occurs at $22\,000$~MCS.  
(c) configuration at 23\,000~MCS,
(d) configuration at 24\,000~MCS,
(e) configuration at 25\,000~MCS,
(f) configuration at 26\,000~MCS, 
}
\label{hyst}
\end{figure}

We also studied the change of the 
configuration at the end of a hysteresis loop, i.e., near the
(pseudo)spinodal that marks the limit of the metastable HS phase.
For this purpose, we decreased the temperature gradually from the HS phase 
in the two-dimensional model with $D=0.4$ ($< D_c \simeq 0.6$). 
The HS state remains as a metastable state beyond the coexistence curve.
However at a certain point, it relaxes quickly to the LS state,  
marking the end of the hysteresis loop near the
(pseudo)spinodal.\cite{UNGE84}
In Fig.~\ref{hyst-m}, we plot the time dependence of the magnetization
as we decrease the temperature in steps by $\Delta T=0.008$ every 40\,000 MCS.
In the figure, the temperature is kept fixed 
at $T=0.1040$, 0.1032, 0.1024 and 0.1016. A rapid change of phase 
takes place at $T=0.1024$. 
In Fig.~\ref{hyst}, we show configurations during this rapid change 
(denoted by circles in Fig.~\ref{hyst-m}).

In contrast to short-range interaction models, in which the phase change
occurs through nucleation and growth of compact critical droplets of the
bulk equilibrium phase,\cite{RTMS} the present system remains
macroscopically uniform during the whole transformation process. 
This is
consistent with the accepted picture of spinodal nucleation in systems
with long-range 
interactions,\cite{UNGE84,MONE92,GORM94,KLEI02,GAGN05,KLEI07} 
where the critical droplet is known to be
extended and highly ramified with a density close to that of the metastable 
phase. It is thus extremely difficult to distinguish from the 
metastable background. 
Growth of this critical droplet occurs by a filling-in of its
``interior," which is seen as the uniform change in the order parameter 
in Fig.~\ref{hyst-m}. 


\subsection{Spin configuration after quench into the low-temperature phase}

\begin{figure}[t]
\centerline{\includegraphics[clip,width=5.5cm] {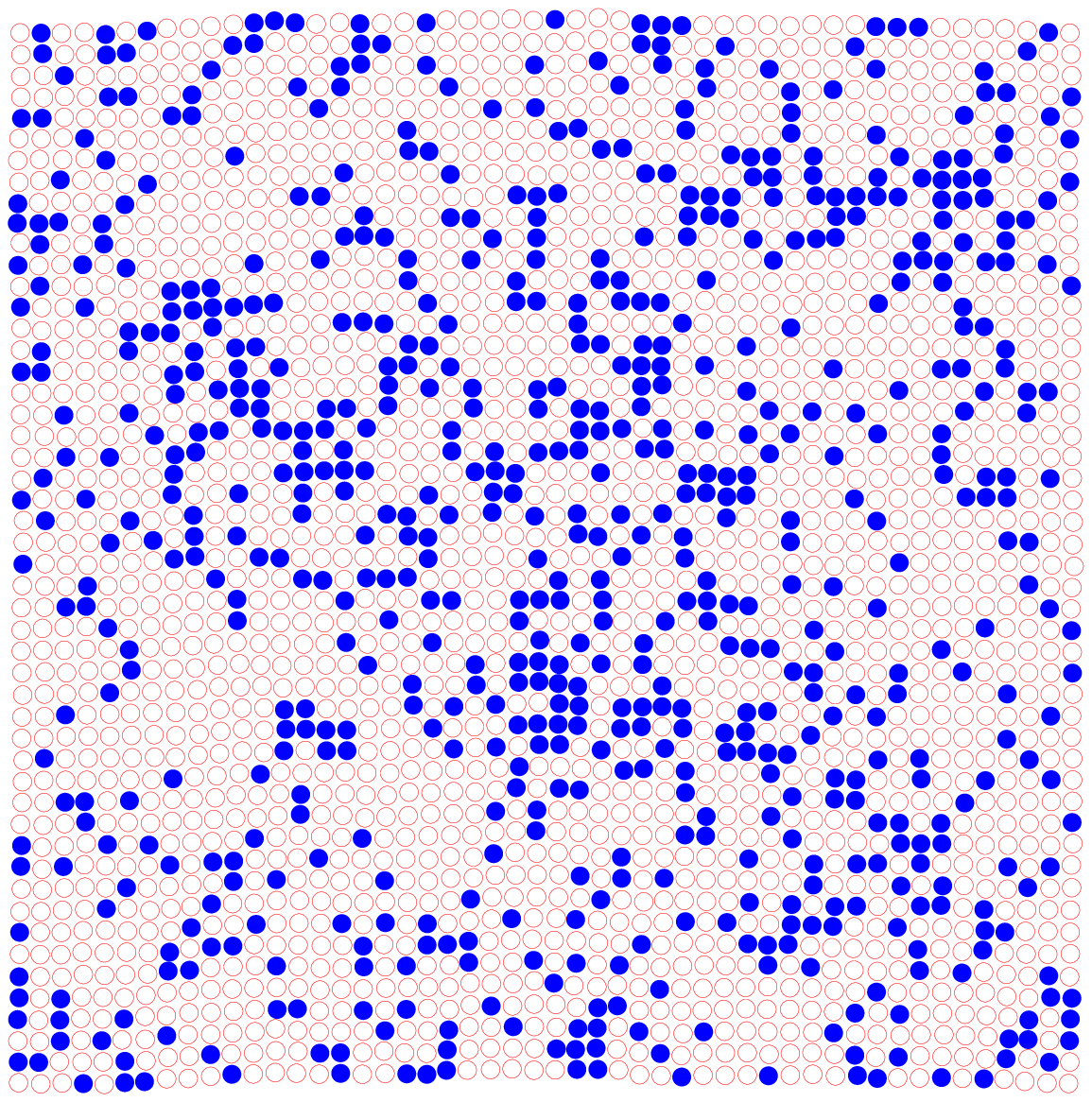} }
\caption{
(Color online)
A spin configuration at 200~MCS after a sudden change of $T$ 
($T=10T_{c}$ to $T=0.8T_{c}$) along the symmetry line, $D=T\ln g$. }
\label{quench}
\end{figure}

In short-range models with non-conserved order parameter,
the cluster size increases proportionally to the square-root of
the elapsed time after a sudden quench from a disordered phase to a 
low-temperature phase.\cite{GUNT83} 
In contrast, the present model does not show such clustering configurations.
In Fig.~\ref{quench}, we show a typical configuration after quenching.
Here we again find no large cluster growth, which
indicates that there is no critical opalescence in the present model.

These processes keeping uniformity can be understood in the following way.
If a large domain exists, it causes a large distortion of the lattice, which is
energetically unfavorable. Thus the system tends to be uniform on large
length scales. This mechanism would be a characteristic of the present 
elastically induced mean-field phase transition. 
Beside the present SC system, there are various 
systems in which elastic interactions play an important role. 
For example, for the martensite transition in metals,\cite{martensite} 
the elastic interaction is important, and 
we expect similar critical behavior there.  

As mentioned previously, as far as a specific material is concerned,
$D$ is given, and the temperature dependence of ordering
is given by the dotted lines in Fig.~2. 
Thus, in most cases the phase transition 
is of first order. In such cases, the $D$ dependence of the 
ordering studied in this paper
is difficult to observe. However, the fact that the system 
is always uniform and 
no clustering occurs should be observable, even in a specific material.
Moreover, by making use of the pressure dependence,\cite{Konishi2007}
we may also observe the critical properties and 
confirm the mean-field universality class.

\section{Summary and Discussion}
\label{sec:DISC}

We studied the critical properties of the elastically induced spin-crossover 
phase transition, finding it to belong to the mean-field universality class.
The temperature dependences of the long-range order and the susceptibility were 
obtained in two- and three-dimensional models, 
and the corresponding critical exponents 
$\beta$ and $\gamma$ were found to be $1/2$ and $1$, respectively, 
in agreement with the mean-field universality class. 
The size- and temperature-dependence of $\langle M^2 \rangle$ converged onto
a scaling function. 
In the analysis of the finite-size scaling, we need critical exponents
for the spin correlations, i.e., $\eta$ and $\nu$. 
We found that the {\it effective\/} values, 
$\eta^*=(4-d)/2$ and $\nu^*=2/d$, are good for the scaling plots,
as has been pointed out in various 
studies of the mean-field universality class.       
We also found that the critical properties of our model agree well with 
the long-range interaction model
(Husimi-Temperley model), in which the spin correlation function is constant 
at large distances. 

We also studied characteristics of the spin configurations 
of the present model with 
effective long-range interactions.
We found that the system does not show configurations with large clusters, 
even following sudden temperature quenches, 
or at the edge of the hysteresis loop near the (pseudo)spinodal.
Thus  critical opalescence 
and conventional nucleation phenomena do not appear in the present model. 
In materials, it is difficult to change $D$ or $g$, but the pressure
dependence of these parameters\cite{Konishi2007} will
enable them to be controlled, 
and we hope that the characteristic behaviors uncovered in this study 
will be found in real experiments in the future.

\section*{Acknowledgments}
\label{sec:ACK}

This work was partially supported by a Grant-in-Aid for Scientific 
Research on Priority Areas
``Physics of new quantum phases in superclean materials" 
(Grant No.\ 17071011) and Grant-in-Aid for Young Scientist (B) from MEXT, 
and also by the Next Generation Super Computer Project, 
Nanoscience Program of MEXT.
Numerical calculations were done on the supercomputer of ISSP. 
This work was also supported by the MST Foundation.
P.A.R.\ gratefully acknowledges hospitality at The University of Tokyo,
as well as useful discussions or correspondence with 
V.~Dobrosavljevic, A.~El-Azab, E.~Luijten, and M.~A.\ Novotny. 
Work at Florida State University was supported by U.S.\ National Science
Foundation Grant No.\ DMR-0444051. 


\section*{Appendix A: Finite size properties of the Husimi-Temperley model}

\begin{figure}[t]
\centerline{\includegraphics[clip,width=6.5cm]{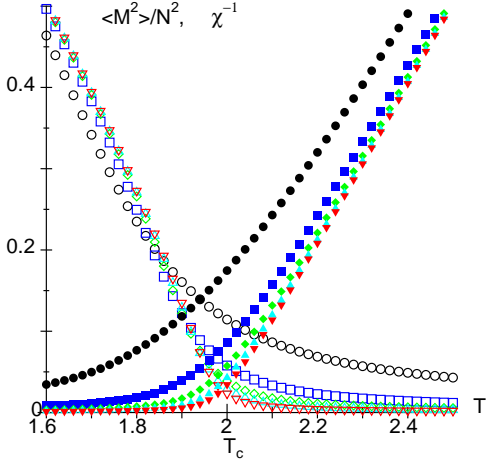} }
\caption{
(Color online) 
Temperature dependence of $\langle M^2\rangle/N^2$ and the inverse susceptibility
$\chi^{-1}=TN/\langle M^2\rangle$ 
for the Husimi-Temperley model. 
The circle, square, diamond, triangle and inverse-triangle denote 
$N=100$, 400, 900, 1600, and 2500, respectively.
The open and closed symbols denote  $\langle M^2\rangle/N$ and $\chi^{-1}$, respectively.
}
\label{fig-HT-M2chi}
\end{figure}

We study the finite-size dependence of
the magnetization of the Husimi-Temperley model 
as a reference of the mean-field type behavior.
The Hamiltonian is given by
\beq
{\cal H}=-{J\over N}\sum_{ij}\sigma_i\sigma_j=-{J\over N}\left(\sum_{i=1}^N\sigma_i\right)^2.
\eeq
Following the standard method, we obtain the partition function:
$$
Z={\rm Tr}e^{-\beta {\cal H}}=e^{{\beta J\over N}\left(\sum_{i=1}^N\sigma_i\right)^2}
$$
$$
={\rm Tr}\int_{-\infty}^{\infty}{dx\over\sqrt{2\pi}}e^{-{1\over 2}x^2+x\sqrt{2\beta J/N}\sum_{i=1}^N\sigma_i}
$$
$$
=
\int_{-\infty}^{\infty}{dx\over\sqrt{2\pi}}e^{-{1\over 2}x^2+N\ln\left[2\cosh\left(x\sqrt{2\beta J/N}
\right)\right]}
$$
\beq
=\int_{-\infty}^{\infty}{\sqrt{N}dz\over\sqrt{2\pi}}e^{-{N\over 2}z^2+N\ln\left[2\cosh\left(z\sqrt{2\beta J}
\right)\right]}
\eeq
If we estimate this integral by the saddle-point method, we obtain the mean-field free energy
\beq
-{\beta F_{\rm MF}\over N}=-{1\over 2}z^2+\ln\left[2\cosh\left(z\sqrt{2\beta J}\right)\right].
\eeq
Here, we obtain the physical quantities of the model for finite values of $N$.
The average of the square of the magnetization is given by
$$
\langle M^2\rangle={{\rm Tr}M^2e^{-\beta {\cal H}} \over {\rm Tr}e^{-\beta {\cal H}}}
=N{\partial\over\partial \beta J}\ln Z
$$
\beq
={
\int_{-\infty}^{\infty}{\sqrt{N}dz\over\sqrt{2\pi}}
N^2{z\over \sqrt{2\beta J}} \tanh\left(z\sqrt{2\beta J}\right)
e^{-{N\over 2}z^2+N\ln\left[2\cosh\left(z\sqrt{2\beta J}
\right)\right]}
\over
\int_{-\infty}^{\infty}{\sqrt{N}dz\over\sqrt{2\pi}}
e^{-{N\over 2}z^2+N\ln\left[2\cosh\left(z\sqrt{2\beta J}
\right)\right]}
},
\eeq
where
\beq
M=\sum_{i=1}^N\sigma_i.
\eeq
The temperature dependence of $\langle M^2\rangle/N^2$ corresponds to the square of the 
spontaneous magnetization $m_{\rm s}$,
\beq
\lim_{N\rightarrow \infty}{\langle M^2\rangle \over N^2}=m_{\rm s}^2,
\eeq
and $k_{\rm B}TN/\langle M^2\rangle$ corresponds to the inverse susceptibility
$\chi^{-1}$ above the critical temperature. We plot the data in Fig.~\ref{fig-HT-M2chi}.
In the present model, the critical temperature is $2J/k_{\rm B}$. Hereafter we
put $J=1$ and $k_{\rm B}=1$.

\begin{figure}[t]
\centerline{\includegraphics[clip,width=6.5cm]{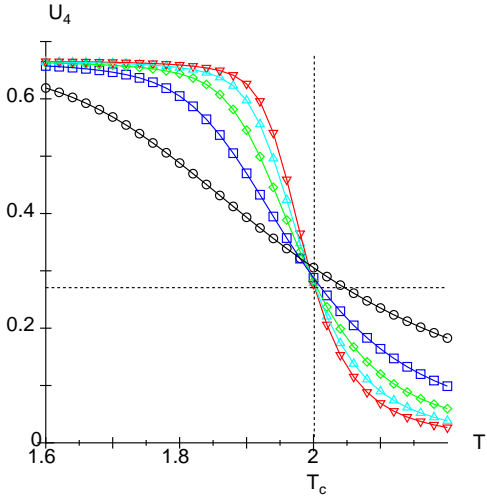} }
\caption{
(Color online)
The Binder plot 
for the Husimi-Temperley model. 
The circle, square, diamond, triangle and inverse-triangle denote 
$N=100,400,900,1600$ and 2500, respectively.
}
\label{fig-HT-binder}
\end{figure}

The Binder plot of this model
is depicted in Fig.~\ref{fig-HT-binder}. 
We find that the data for large $N$ show a good crossing at $T_{c}$.

\begin{figure}[t]
\centerline{\includegraphics[clip,width=6.5cm]{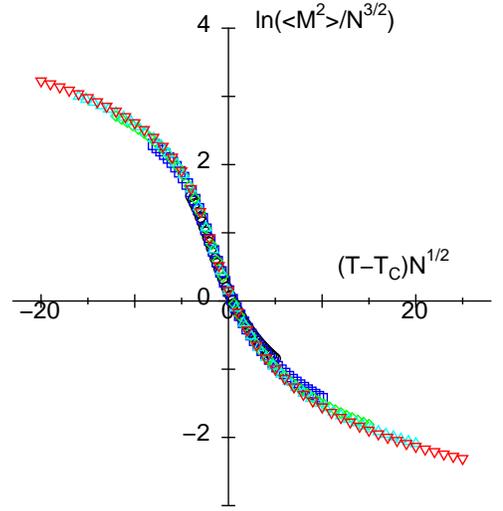} }
\caption{
(Color online)
The scaling plot
of $\langle M^2 \rangle$ for the Husimi-Temperley model. 
The circle, square, diamond, triangle and inverse-triangle denote 
$N=100$, 400, 900, 1600, and 2500, respectively.}
\label{fig-HT-scaling}
\end{figure}

The following size dependences are easily obtained:
\beq 
\langle M^2 \rangle \propto 
\left\{ \begin{array}{cc}
N^2\quad & {T<T_{c}},\\
N^{3/2}  & \quad {T=T_{c}},\\
N\quad   & {T>T_{c}}.
\end{array}\right.
\label{MFsizedep}
\eeq
The size- and temperature dependence of $\langle M^2\rangle$ is found to converge 
in the standard finite size scaling plot as depicted in Fig.~\ref{fig-HT-scaling}.
The values of the correlation functions at 
large separations 
in Eq.~(\ref{corpara}) and  Eq.~(\ref{cortc}) 
correspond to the above size dependences.

These figures qualitatively agree well with those for the model of the elastic interaction
mediated spin-crossover materials.

\begin{figure}[t]
\centerline{\includegraphics[clip,width=6.5cm]{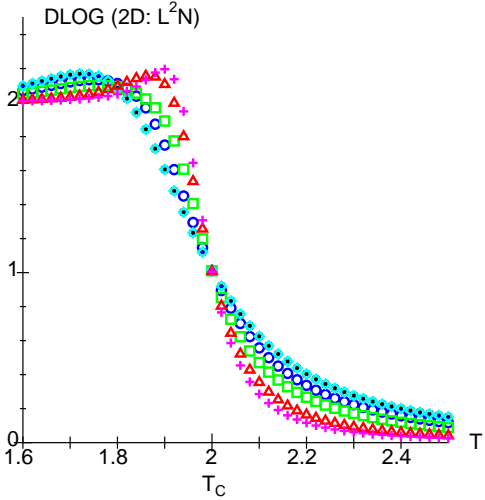} }
\caption{The phenomenological renormalization plot for the
Husimi-Temperley model. Data for 
$(L,L')=$ (10,20), (10,30), (10,50), (20,30), (20,50), and (30,50)
are plotted by cross, circle, square, diamond, triangle, and 
plus, respectively.}
\label{fig-DLOG}
\end{figure}

In Fig.~\ref{fig-DLOG} we 
plot the phenomenological scaling plot of the present data
\beq
{\rm DLOG}={\ln{\langle M^2\rangle_L/ \langle M^2\rangle_L'}\over \ln{L/L'}},
\eeq
for various sets of $(L,L')$. Here, we define $L=N^{1/2}$. 
In general, if we use a definition  $L=N^{1/d}$, DLOG becomes DLOG$\times (d/2)$.

\section*{Appendix B: Finite-size scaling of the mean-field model}

In this Appendix we summarize the finite-size scaling properties expected
in a spatially extended system with mean-field behavior, which agree
with those observed numerically in this paper.  

A $d$-dimensional $\phi^4$ lattice field theory with interaction range 
$\mathcal{R}$ can be defined by the Ginzburg-Landau Hamiltonian
in reciprocal space, 
\begin{eqnarray}
\frac{\tilde{{\mathcal H}}(\phi_{\bf k})}{k_{\rm B}T} 
&=& \frac{1}{2} \sum_{\bf k} \left[ (\mathcal{R} k)^\sigma - t \right] 
    \phi_{\bf k}\phi_{-{\bf k}} \nonumber\\
&&  + \frac{u}{4N} \sum_{{\bf k}_1,{\bf k}_2,{\bf k}_3}
    \phi_{{\bf k}_1} \phi_{{\bf k}_2} \phi_{{\bf k}_3}    
    \phi_{-({\bf k}_1+{\bf k}_2+{\bf k}_3)} \nonumber\\
&&  - h\sqrt{\frac{N}{2}} \phi_{{\bf k}={\bf 0}}
\;,
\label{eq:GL}
\end{eqnarray}
where $N=L^d$ is the number of lattice points, $t = (T-T_c)/T_c$, 
and $h$ is an applied magnetic field. 
For $\sigma=2$ and constant $\mathcal{R}$, this model has local
interactions and upper critical dimension $d_u = 4$. For $d > 4$ it
has classical mean-field critical exponents, for $d=4$ it has
mean-field exponents with logarithmic corrections, and for $d<4$ it
has nontrivial critical exponents corresponding to the $d$-dimensional Ising
universality class.\cite{Luijten-Bloete} 

{}For $d$ below four there are
several ways the model can be modified to show mean-field critical
behavior. One is to keep $\sigma=2$ fixed and let 
$\mathcal{R} \rightarrow \infty$ while using a scaling ansatz equivalent to a
Ginzburg criterion,\cite{Rikvold1993} as is often done in
studies of crossover scaling.\cite{LUIJ98} In this limit of infinitely
weak, infinitely long-ranged interactions, the model reduces to the
Husimi-Temperley model discussed in Appendix A. 
However, the method most relevant to elastic 
systems\cite{TEOD82,PEYL99,PEYL03,UEMU01} is probably
to increase the interaction range by modifying $\sigma$.\cite{Luijten-Bloete} 
{}For $0 \le \sigma < d/2$, this lowers the upper critical dimension to 
\begin{equation}
d_u(\sigma) = 2 \sigma 
\label{eq:du}
\end{equation}
and leads to classical mean-field critical behavior for $d > d_u(\sigma)$.  
(As for $\sigma=2$, 
classical exponents with logarithmic corrections are found for 
$d = d_u(\sigma)$.)

>From the terms corresponding to ${\bf k}={\bf 0}$ in 
Eq.~(\ref{eq:GL}), one gets the standard mean-field 
critical exponents for a spatially uniform system, 
\begin{equation}
\beta = 1/2
\label{eq:bet}
\end{equation}
for the temperature dependence of the order parameter, 
$\overline{\phi} \propto |t|^\beta$ for $t \le 0$, 
\begin{equation}
\delta = 3
\label{eq:delt}
\end{equation}
for its field dependence, 
$\overline{\phi} \propto |h|^{1/\delta}$ for $t=0$, and  
\begin{equation}
\gamma = \gamma' = 1 
\label{eq:gam}
\end{equation}
for the corresponding susceptibility, 
$\chi = \partial \overline{\phi} / \partial h \propto |t|^{- \gamma}$. 

Spatial fluctuations are governed by the $k^\sigma$ term. In the
Gaussian approximation this yields 
\begin{equation}
\nu = 1/\sigma 
\label{eq:nuG}
\end{equation}
for the correlation length, 
$\xi \propto |t|^{-\nu}$,
and 
\begin{equation}
\eta = 2 - \sigma 
\label{eq:etaG}
\end{equation}
for the spin correlation function, 
$c(r) \propto \exp [-r/\xi] r^{-(d-2+\eta)}$.\cite{GOLD92}
However, renormalization of the ``dangerous irrelevant variable" $u$
that multiplies the fourth-order term in Eq.~(\ref{eq:GL})\cite{PRIV83,GOLD92} 
causes the fluctuations on large length scales comparable to the linear
system size $L$ instead to be governed by the $\sigma$-independent {\it
effective\/} exponents,\cite{PRIV83,Luijten-Bloete} 
\begin{equation}
\nu^* = 2/d 
\label{eq:nuE}
\end{equation}
and 
\begin{equation}
\eta^* = 2 - d/2 
\label{eq:etaE}\;.
\end{equation}

Using these effective exponents in standard finite-size scaling
relations,\cite{PRIV84,PRIV90},
one obtains the following scaling relation,
\begin{eqnarray}
\langle M^2 \rangle 
&\propto& 
L^{2d} \left[ L^{-2 \beta/\nu^*} + L^{2-\eta^*} \right] 
\mathcal{M}^2(tL^{1/\nu^*}) 
\nonumber\\
&=& 
L^{3d/2} \mathcal{M}^2(tL^{d/2}) \;,
\label{eq:M2s}
\end{eqnarray}
where the scaling relation $\gamma/\nu^* = 2 - \eta^*$ has been used, and 
$\mathcal{M}^2(x)$ is a scaling function.\cite{binder1985} 
The Binder cumulant $U_4$ also
becomes a scaling function of $x=tL^{d/2}$, ranging from 2/3 for $x \ll
0$ to 0 for $x \gg 0$ with the fixed-point value of Eq.~(\ref{eq:U4star}) at
$x=0$. Similarly, the phenomenological renormalization plot obtained from
Eq.~(\ref{eq:DLOG}) will go from $2d$ for $t \ll 0$ through 
$d+2-\eta^* = 2(d-\beta/\nu^*) = 3d/2$ at $t=0$, to $-d$ for $t \gg 0$. 
The spin correlation function at $r \sim L$ takes the value 
$c_0 \propto L^{-(d-2+\eta^*)} = L^{-d/2}$. For $r \ll L$, on the other
hand, the behavior is expected to be governed by the Gaussian exponent
$\eta$. However, much larger systems than
the ones studied here would be needed to detect this 
behavior, which would enable one to measure the value of 
$\sigma$.\cite{Luijten-Bloete}

We finally note two interesting aspects of these mean-field finite-size scaling
relations. First, by making the replacement $L^d = N$, it is easy to see
that they all become independent of $d$, as long as $d > d_u(\sigma)$. 
Second, we note that the Gaussian exponents can be used as in ordinary
finite-size scaling theory, provided that $L$ is replaced by the
{\it modified system size\/}, $L^{d/d_u(\sigma)}$.\cite{Jones}

The details of the effective long-range interactions
introduced by the elastic degrees of freedom in the present system 
are not known, except for $d=1$. In this case it has been shown rigorously
that the model can be mapped onto an Ising chain with nearest-neighbor
ferromagnetic interactions,\cite{KBO2} and thus it exhibits no phase
transition at nonzero temperatures. 
Much work has been devoted over the years
to the interactions between defects in
three-dimensional elastic solids, and it is generally argued that the
dominant long-range interactions are dipole-dipole interactions $\sim 1/r^3$ 
that can be attractive or repulsive depending on the
relative orientations of the dipoles.\cite{TEOD82}  
The case of $d=2$ has been much less studied, and relevant works are
much more recent. Dimensional analysis
indicates that dipole-dipole interactions $\sim 1/r^2$ should be present
unless forbidden by symmetry.\cite{PEYL99,PEYL03,UEMU01}
Although the effects of distortions in the present model are not
identical to those in the classic elastic media for which these results
were obtained, we do not think it is unreasonable to assume that
the elastically mediated interactions in our model are of such long range type.  
This then would lead to $\sigma =0$ and consequently 
$d > d_u=0$ for both cases, so that classical mean-field critical
behavior would indeed be expected. However, a weaker condition on
$\sigma$, which still would lead to mean-field critical behavior, 
is obtained by simply requiring $d > d_u(\sigma)$, leading to $0 \le
\sigma < d/2$ or, equivalently, interactions $\sim 1/r^{d+\sigma}$ with 
$d \le d+\sigma < 3d/2$. We also note that the mechanism involving
$\sigma =2$ and $\mathcal{R} \rightarrow \infty$ leads to the same
effective exponents as the variable-$\sigma$
mechanism,\cite{Rikvold1993,Jones}
and so it would also be consistent with our numerical data. 
Which (if any) of these mechanisms best describes the long-range
interactions that cause the mean-field
critical behavior observed in the model studied here, remains an interesting
question for future research.


\begin{thebibliography}{99}
\bibitem{update}
In our model, the update procedure is symmetric for HS and LS.
The effect of the volume change is taken into account in the transition probability.
Thus, the symmetry-breaking field is given as 
$H=\frac{1}{2}(D-k_{\rm B}T_c\ln g)=0$.
\bibitem{shotrange-order}
In the vertical (or horizontal) direction, the present model has 
antiferromagnetic short-range order, which will be analyzed 
in detail elsewhere.
\end{thebibliography}
\end{document}